\documentclass[11pt]{article}
\usepackage{a4,bm,epsfig}
\usepackage{ifthen,array}
\newcommand{\ams}{\usepackage{amsfonts,amssymb,amsmath}}

\ams
\allowdisplaybreaks[3]
\newlength{\textwidthorig}
\newlength{\oddsidemarginorig}
\newlength{\textheightorig}
\newlength{\topmarginorig}
\def\seitenlaengenabsolut#1 #2 #3 #4 {\setlength{\textwidth}{#1}
                                      \setlength{\oddsidemargin}{#2}
                                      \setlength{\textheight}{#3}
                                      \setlength{\topmargin}{#4}}
\def\seitenlaengenrelzustandard#1 #2 #3 #4 {\setlength{\textwidth}{\textwidthorig+#1}
                                            \setlength{\oddsidemargin}{\oddsidemarginorig+#2}
                                            \setlength{\textheight}{\textheightorig+#3}
                                            \setlength{\topmargin}{\topmarginorig+#4}}
\def\seitenlaengenrelzuvorher#1 #2 #3 #4 {\addtolength{\textwidth}{#1}
                                          \addtolength{\oddsidemargin}{#2}
                                          \addtolength{\textheight}{#3}
                                          \addtolength{\topmargin}{#4}}
\newcommand{\standardseite}{\seitenlaengenrelzuvorher2.2cm -0.8cm 1.8cm -1.5cm }   %
\standardseite

\newlength{\laengespatium}

\newcommand{\nach}{\longrightarrow}      

\newcommand{\auf}{\longmapsto}           
\newcommand{\txtauf}[1]{\auf}            
\newcommand{\impliz}{\Longrightarrow}    
\newcommand{\aequ}{\Longleftrightarrow}  
 
\newcommand{\invimpliz}{\Longleftarrow}  
\newcommand{\gegen}{\rightarrow}         
\newcommand{\konvhoch}{\uparrow}         
\newcommand{\iso}{\cong}                 
\newcommand{\kong}{\ident}               
\newcommand{\ident}{\equiv}              
\newcommand{\teilmenge}{\subseteq}       
\newcommand{\obermenge}{\supseteq}       
\newcommand{\aeqrel}{\sim}               
\newcommand{\nichtin}{\not\in}

\newcommand{\leeremenge}{\varnothing}    
\newcommand{\tensor}{\otimes}            
\newcommand{\kreuz}{\times}              

\newcommand{\einschr}[1]{{}\arrowvert_{#1}}      

\newcommand{\bigtensor}{\bigotimes}      
\newcommand{\betraganpass}[1]%
           {\left| #1 \right|}           
\newcommand{\bigbetrag}[1]%
           {\bigl|{#1}\bigr|}            
\newcommand{\betrag}[1]%
           {|{#1}|}                      
\newcommand{\betragnichtanpass}[1]%
           {\mid #1 \mid}                
\newcommand{\norm}[1]%
           {{}{\parallel}#1{\parallel}{}}      
\newcommand{\erww}[1]%
           {\langle #1 \rangle}          
\newcommand{\skalprod}[2]%
           {\langle #1,#2 \rangle}       
\newcommand{\quer}{\overline}            
\newcommand{\dach}{\widehat}             
\newcommand{\inv}[1]{\frac{1}{#1}}       
\newcommand{\im}{\text{im\;}}                          
\newcommand{\tr}{\text{tr}}                           
\newcommand{\elanz}{\#}                                
\newcommand{\Hom}{\text{Hom}}                          
\newcommand{\End}{\text{End}}                          
\newcommand{\dd}{\text{d}}                             
\newcommand{\field}[1]{\mathbb{#1}}                    
\newcommand{\C}{{\field{C}}}                           
\newcommand{\N}{{\field{N}}}                           
\newcommand{\R}{{\field{R}}}                           
\newcommand{\rnkl}[2]{\raisebox{-0.4ex}{$#1$}%
\raisebox{-0.12ex}{{\large$\setminus$}}\,#2}   
\newcommand{\agb}{{\overline{{\cal A}/{\cal G}}}}      
\newcommand{\agbfact}[1][]{\text{$\agb/\!\aeqrel$}}    
\newcommand{\Ab}{{\overline{{\cal A}}}}                
\newcommand{\A}{{\cal A}}                              
\newcommand{\Gb}{{\overline{{\cal G}}}}                

\newcommand{\qa}{{\quer{A}}}                           

\newcommand{\holgr}{{\mathbf H}}                       
\newcommand{\bz}{{\mathbf B}}                          

\newcommand{\gross}[1]{{\boldsymbol #1}}               
\newcommand{\gc}{\gross{\gamma}}                       
\newcommand{\Pf}{{\cal P}}                             
\newcommand{\KG}[1]{\Pf_{#1}}                          
\newcommand{\BB}{\uparrow\uparrow}                     
\newcommand{\EB}{\downarrow\uparrow}                   
\newcommand{\BE}{\uparrow\downarrow}                   
\newcommand{\EE}{\downarrow\downarrow}                 
\newcommand{\notBB}{\mbox{${}\BB{}$\hspace*{-2.8ex}\rule[0.40ex]%
                    {2ex}{0.4pt}\hspace*{0.8ex}}}      
\newcommand{\hyph}{\upsilon}                           
\newcommand{\Haar}{{\text{Haar}}}                      
\newcommand{\LG}{{\mathbf{G}}}                         
\newcommand{\aeqrelzush}[1][]{\sim}                    

\newcommand{\nklza}[1][]{\ifthenelse{\equal{#1}{}}     
                                    {\rnkl{Z(\holgr_\qa)}{\LG}}        
                                   {\rnkl{Z(\holgr_{#1})}{\LG}}}       
\newcommand{\nkla}[1][]{\ifthenelse{\equal{#1}{}}      
                                    {\rnkl{\bz(\qa)}{\Gb}}        
                                    {\rnkl{\bz(#1)}{\Gb}}}       

\newcommand{\darst}{{\phi}}                            

\newcommand{\YM}{{\text{YM}}}                          

\newcommand{\ymwirk}[1][]{\ifthenelse{\equal{#1}{}}{S_{\YM}}{S_{\YM,#1}}}

\newcommand{\bmat}{\begin{pmatrix}}
\newcommand{\emat}{\end{pmatrix}}

\newcommand{\ListNullAbstaende}{\setlength{\topsep}{0pt}%
                                \setlength{\parskip}{0pt}%
                                \setlength{\partopsep}{0pt}%
                                \setlength{\itemsep}{0pt}%
                                \setlength{\parsep}{0pt}}
\newcommand{\ListNurAnstrichAbstand}{\setlength{\topsep}{0pt}%
                                     \setlength{\parskip}{0pt}%
                                     \setlength{\partopsep}{0pt}%
                                     \setlength{\parsep}{0pt}}
\newenvironment{StandardListe}[2]%
               {\begin{list}%
                      {#1}%
                      {\settowidth{\leftmargin}{M#1}%
                       \settowidth{\labelwidth}{#1}%
                       \settowidth{\labelsep}{M}%
                       #2%
                      }%
                }%
               {\end{list}}%
\newenvironment{EinfachListe}[1]%
               {\begin{StandardListe}{#1}{\ListNullAbstaende}}%
               {\end{StandardListe}}%
               {\begin{StandardListe}{#1}{\ListNurAnstrichAbstand}}%
               {\end{StandardListe}}%
\newcommand{\labelsatz}[1]{#1}
\newcounter{listennr}                      %
\newlength{\hilfslaenge}
\newlength{\stdlabellaenge}
\newlength{\maximum}
\newcommand{\stdlabel}{}
\newcommand{\Maximum}{}
\newcommand{\iitem}[1][]{\ifthenelse{\equal{#1}{}}%
                           {\item \setlength{\hilfslaenge}{\stdlabellaenge}}%
                           {\item[\labelsatz{#1}\hfill]%
                            \settowidth{\hilfslaenge}{\labelsatz{#1}}}%
                         \ifthenelse{\lengthtest{\maximum < \hilfslaenge}}%
                           {\setlength{\maximum}{\hilfslaenge}%
                            \ifthenelse{\equal{#1}{}}%
                               {\renewcommand{\Maximum}{\stdlabel}}%
                               {\renewcommand{\Maximum}{#1}}}%
                           {}%
                      }      
\makeatletter
\newenvironment{AutoLabelLaengenListe}[2][]%
               {\begin{list}%
                      {\labelsatz{#1}\hfill}%
                      {\stepcounter{listennr}%
                       \settowidth{\leftmargin}{M\labelsatz{\ref{listnr\arabic{listennr}}}}%
                       \settowidth{\labelwidth}{\labelsatz{\ref{listnr\arabic{listennr}}}}%
                       \settowidth{\labelsep}{M}%
                       \settowidth{\stdlabellaenge}{\labelsatz{#1}}%
                       \renewcommand{\stdlabel}{#1}%
                       #2%
                       \renewcommand{\Maximum}{}%
                      }%
                }%
               {\renewcommand{\@currentlabel}{\Maximum}%
                \label{listnr\arabic{listennr}}%
                \end{list}%
                }%
\makeatother
\newenvironment{StandardEinrueckung}[2]%
               {\begin{list}%
                      {#1}%
                      {\settowidth{\leftmargin}{M#1}%
                       \settowidth{\labelwidth}{#1}%
                       \settowidth{\labelsep}{M}%
                       #2%
                      }%
                \item}%
               {\end{list}}%
\newenvironment{Einrueckungpur}[1]%
               {\begin{StandardEinrueckung}{#1}{\ListNullAbstaende}}%
               {\end{StandardEinrueckung}}%
\newenvironment{Einrueckung}[1]%
               {\begin{StandardEinrueckung}{#1}{\setlength{\parsep}{0pt}}}%
               {\end{StandardEinrueckung}}%
\newcommand{\EineZeileGleichung}[2][0.0ex]
           {
            
            \vspace{#1} 
            \noindent
            \hspace*{\fill}
            $\displaystyle{#2}$
            \hspace*{\fill}

            \vspace{#1} 
            
           }

\makeatletter
\newcommand{\EineNumZeileGleichung}[2][0.5ex]
           {
            
            \vspace{#1} 
            \noindent
            \stepcounter{equation}
            \renewcommand{\@currentlabel}{\arabic{equation}}%
            \phantom{(\arabic{equation})}\hspace*{\fill}
            $\displaystyle{#2}$
            \hspace*{\fill}
            (\arabic{equation})

            \vspace{#1} 
            
           }
\makeatother
\makeatletter
\newcommand{\EineErwNumZeileGleichung}[2][0.5ex]
           {
            
            \vspace{#1} 
            \noindent
            \stepcounter{equation}
            \renewcommand{\@currentlabel}{\arabic{equation}}%
            \phantom{(\arabic{equation})}\hspace*{\fill}
            #2 %
            \hspace*{\fill}
            (\arabic{equation})

            \vspace{#1} 
            
           }
\makeatother
\newcommand{\breitrel}[1]{\hspace*{\tabcolsep} #1 \hspace*{\tabcolsep}}
\newlength{\abstaug}              %
\newenvironment{AllgUnnumGleichung}[2][1.0ex]
               {
  
                \setlength{\abstaug}{#1}
                \vspace{\abstaug}
                \hspace*{\fill}
                $\begin{array}[t]{#2}
                }%
               {\end{array}$
                \hspace*{\fill}
  
                \vspace{\abstaug}

                }%
\newenvironment{AllgNumGleichung}[2][0.0ex]
               {
  
                \setlength{\abstaug}{#1}
                \vspace{\abstaug}
                $\begin{tabular*}{\textwidth}[t]{#2}
                }%
               {\end{tabular*}$

                \vspace{\abstaug}

               }%
\newenvironment{StandardUnnumGleichungKlein}[1][0ex]
               {\renewcommand{\s}{\\[#1] }%
                \begin{AllgUnnumGleichung}{rcl}}%
               {\end{AllgUnnumGleichung}}%
\newcommand{\s}{\\[0ex] }             %
\newenvironment{StandardUnnumGleichung}[1][0ex]%
               {\renewcommand{\s}{\\[#1] }%
                \begin{AllgUnnumGleichung}{>{\displaystyle}rc>{\displaystyle}l}}%
               {\end{AllgUnnumGleichung}}%
\newenvironment{XrelYZNumGleichung}[1][0ex]
               {\renewcommand{\s}{\\[#1] }%
                \begin{AllgNumGleichung}{rcll}}%
               {\end{AllgNumGleichung}}%
\newcommand{\erl}[1]{\hfill\mbox{\hspace*{1.5em}\small (#1)}}

\newcommand{\erllang}[2][0.5\textwidth]%
              {\hfill\hspace*{1.5em}%
               \begin{minipage}[t]{#1}{\small%
                          \begin{list}{(}{\ListNullAbstaende%
                                          \settowidth{\leftmargin}{(}%
                                          \settowidth{\labelwidth}{(}%
                                          \settowidth{\labelsep}{}%
                                         }%
                          \item#2)%
                          \end{list}}%
               \end{minipage}\\[-0.9ex]
              }%
\newcommand{\DefBemUmgeb}[1]%
           {\newenvironment{#1}[1][]%
                           {\begin{Einrueckung}{{\bf #1}}%
                            \ifx##1\empty\else{{\bf ##1}
                            
                                                        }\fi%
                            }%
                           {\end{Einrueckung}}}
\newcommand{\DefSBemUmgeb}[2]
           {\newenvironment{#1}[1][]%
                           {\begin{Einrueckung}{{\bf #2}}%
                            \ifx##1\empty\else{{\bf ##1}
                            
                                                        }\fi%
                            }%
                           {\end{Einrueckung}}}
\makeatletter
\newcommand{\DefBspUmgeb}[3]
           {\newcounter{#2}[#3]%
            \newenvironment{#1}[1][]%
                           {\stepcounter{#2}%
                            \renewcommand{\ZaehlerMarke}{\arabic{#2}}%
                            \renewcommand{\Einzugsname}{{\bf #1 \ZaehlerMarke}}%
                            \begin{Einrueckung}{\Einzugsname}
                            \ifx##1\empty\else{{\bf ##1}\\}\fi%
                            \renewcommand{\@currentlabel}{\ZaehlerMarke}%
                            }%
                           {\end{Einrueckung}}}
\makeatother
\newcommand{\ZaehlerbisEbene}{section}
\newcommand{\Ebenea}{section}
\newcommand{\Ebeneb}{subsection}

\newcommand{\Abschnittnummer}{%
            \ifx\ZaehlerbisEbene\Ebenea{\arabic{section}}%
             \else{%
              \ifx\ZaehlerbisEbene\Ebeneb{\arabic{section}.\arabic{subsection}}%
               \else{\arabic{section}.\arabic{subsection}.\arabic{subsubsection}}%
              \fi}%
            \fi}     
\newcommand{\Abschnittnummerpunkt}{\Abschnittnummer.}     
\newcommand{\Einzugsname}{}
\newcommand{\ZaehlerMarke}{}
\makeatletter
\newcommand{\DefThmUmgeb}[3]%
           {\newcounter{#1}[#3]%
            \newenvironment{#1}[1][]%
                           {\stepcounter{#2}%
                            \setcounter{#1}{\value{#2}}%
                            \renewcommand{\ZaehlerMarke}{\Abschnittnummerpunkt\arabic{#1}}%
                            \renewcommand{\Einzugsname}{{\bf #1 \ZaehlerMarke}}%
                            \begin{Einrueckung}{\Einzugsname}
                            \ifx##1\empty\else{{\bf ##1}
                            
                                                        }\fi%
                            \renewcommand{\@currentlabel}{\ZaehlerMarke}%
                            }%
                           {\end{Einrueckung}}}
\makeatother
\makeatletter
\newcommand{\DefSThmUmgeb}[4]%
           {\newcounter{#1}[#3]%
            \newenvironment{#1}[1][]%
                           {\stepcounter{#2}%
                            \setcounter{#1}{\value{#2}}%
                            \renewcommand{\ZaehlerMarke}{\Abschnittnummerpunkt\arabic{#1}}%
                            \renewcommand{\Einzugsname}{{\bf #4 \ZaehlerMarke}}
                            \begin{Einrueckung}{\Einzugsname}
                            \ifx##1\empty\else{{\bf ##1}

                                                        }\fi%
                            \renewcommand{\@currentlabel}{\ZaehlerMarke}%
                            }%
                           {\end{Einrueckung}}}
\makeatother
\makeatletter
\newcommand{\DefUnterNumThmUmgeb}[5]%
           {\newcounter{#1}[#3]%
            \newcounter{#4}%
            \newenvironment{#1}[1][]%
                           {\ifx##1\empty\else{\stepcounter{#2}\setcounter{#4}{0}}\fi%
                            \stepcounter{#4}%
                            \setcounter{#1}{\value{#2}}%
                            \renewcommand{\ZaehlerMarke}{\Abschnittnummerpunkt\arabic{#1}\alph{#4}}%
                            \renewcommand{\Einzugsname}{{\bf #5 \ZaehlerMarke}}
                            \begin{Einrueckung}{\Einzugsname}
                            \renewcommand{\@currentlabel}{\ZaehlerMarke}%
                            }%
                           {\end{Einrueckung}}}
\makeatother
\newenvironment{Beweis}[1][]%
               {\begin{Einrueckung}{{\bf Beweis}}%
                \ifx#1\empty\else{{\bf #1}

                                            }\fi%
                }%
               {\end{Einrueckung}%
                }%
\newenvironment{Proof}[1][]%
               {\begin{Einrueckung}{{\bf Proof}}%
                \ifx#1\empty\else{{\bf #1}

                                            }\fi%
                }%
               {\end{Einrueckung}%
                }%
               {\begin{Einrueckung}{{\bf \glqq Beweis\grqq}}%
                \ifx#1\empty\else{{\bf #1}
                
                                            }\fi%
                }%
               {\end{Einrueckung}%
                }%
               {\begin{Einrueckung}{{\bf Begr"undung}}%
                \ifx#1\empty\else{{\bf #1}
                
                                            }\fi%
                }%
               {\end{Einrueckung}%
                }%
\newenvironment{Hinrichtung}%
               {\begin{Einrueckungpur}{$\impliz$}}%
               {\end{Einrueckungpur}}%
\newenvironment{Rueckrichtung}%
               {\begin{Einrueckungpur}{$\invimpliz$}}%
               {\end{Einrueckungpur}}%
               {\begin{Einrueckungpur}{\glqq$\teilmenge$\grqq}}%
               {\end{Einrueckungpur}}%
               {\begin{Einrueckungpur}{\glqq$\obermenge$\grqq}}%
               {\end{Einrueckungpur}}%
\newenvironment{SubSet}%
               {\begin{Einrueckungpur}{"$\teilmenge$"}}%
               {\end{Einrueckungpur}}%
\newenvironment{SuperSet}%
               {\begin{Einrueckungpur}{"$\obermenge$"}}%
               {\end{Einrueckungpur}}%
\newcommand{\qed}{\nopagebreak\hspace*{2em}\hspace*{\fill}{\bf qed}}
\newcommand{\ARabic}{\arabic}
\newcommand{\Nummerntypa}{\arabic}   
\newcommand{\Nummerntypb}{\alph}
\newcommand{\Nummerntypc}{\roman}
\newcommand{\Nummerntypd}{\Alph}

\newcommand{\Nra}{\Nummerntypa{Nummera}}            %
\newcommand{\Nrb}{\Nummerntypb{Nummerb}}            %
\newcommand{\Nrc}{\Nummerntypc{Nummerc}}                
\newcommand{\Nrd}{\Nummerntypd{Nummerd}}                
\newcommand{\ZeichenzuNrTyp}[1]%
           {\ifx#1\ARabic {.}\else{)}%
                  \fi}                              %
\newcommand{\NrZeicha}{\ZeichenzuNrTyp{\Nummerntypa}}
\newcommand{\NrZeichb}{\ZeichenzuNrTyp{\Nummerntypb}}
\newcommand{\NrZeichc}{\ZeichenzuNrTyp{\Nummerntypc}}
\newcommand{\NrZeichd}{\ZeichenzuNrTyp{\Nummerntypd}}
\newcommand{\ListMarkea}%
           {\Nra\NrZeicha}
\newcommand{\ListMarkeb}%
           {\Nra\NrZeicha\Nrb\NrZeichb}
\newcommand{\ListMarkec}%
           {\Nra\NrZeicha\Nrb\NrZeichb\Nrc\NrZeichc}
\newcommand{\ListMarked}%
           {\Nra\NrZeicha\Nrb\NrZeichb\Nrc\NrZeichc\Nrd\NrZeichd}
\newcommand{\Anfangszeichen}{}
\newcommand{\Anfangspunkt}{}
\newcounter{Schachtelebene}
\newcounter{Hilfszaehler}
\newcommand{\Hilfsbefehl}{}
\newcommand{\Schachtelebene}{\alph{Schachtelebene}}
\makeatletter
\newenvironment{AllgNumerierteListe}[2][]
               {\addtocounter{Schachtelebene}{1}%
		\setcounter{Hilfszaehler}{#2}%
                \renewcommand{\Anfangszeichen}%
                             {\renewcommand{\Hilfsbefehl}{\csname Nummerntyp\Schachtelebene \endcsname}%
                              \Hilfsbefehl{Hilfszaehler}}%
                \renewcommand{\Anfangspunkt}%
                             {\csname NrZeich\Schachtelebene \endcsname}%
                \begin{list}%
                      {\stepcounter{Nummer\Schachtelebene}%
                       \csname Nr\Schachtelebene \endcsname
                       \csname NrZeich\Schachtelebene \endcsname
                       }%
                      {\settowidth{\leftmargin}{M\Anfangszeichen\Anfangspunkt}%
                       \settowidth{\labelwidth}{\Anfangszeichen\Anfangspunkt}%
                       \settowidth{\labelsep}{M}%
                       \setlength{\topsep}{0pt}%
                       \setlength{\parskip}{0pt}%
                       \setlength{\partopsep}{0pt}%
                       \setlength{\itemsep}{0pt}%
                       \setlength{\parsep}{0pt}%
                      }%
                \renewcommand{\@currentlabel}{\csname ListMarke\Schachtelebene \endcsname}%
                }%
               {\ifthenelse{\equal{}{}}{\setcounter{Nummer\Schachtelebene}{0}}{}
                \addtocounter{Schachtelebene}{-1}%
                \end{list}}
\makeatother
\newenvironment{NumerierteListe}[1]
               {\begin{AllgNumerierteListe}{#1}}
               {\end{AllgNumerierteListe}}
\newenvironment{WeiterNumerierteListe}[1]
               {\begin{AllgNumerierteListe}[Weiter]{#1}}
               {\end{AllgNumerierteListe}}

\newcommand{\UnnumAnfangszeichen}{}
\newcounter{UnnumSchachtelebene}
\newcommand{\UnnumSchachtelebene}{\alph{UnnumSchachtelebene}}
\makeatletter
\newenvironment{UnnumerierteListe}%
               {\addtocounter{UnnumSchachtelebene}{1}%
                \renewcommand{\UnnumAnfangszeichen}%
                             {\csname UnnumZeich\UnnumSchachtelebene \endcsname}%
                \begin{list}%
                      {\UnnumAnfangszeichen}%
                      {\settowidth{\leftmargin}{M\UnnumAnfangszeichen}%
                       \settowidth{\labelwidth}{\UnnumAnfangszeichen}%
                       \settowidth{\labelsep}{M}%
                       \setlength{\topsep}{0pt}%
                       \setlength{\parskip}{0pt}%
                       \setlength{\partopsep}{0pt}%
                       \setlength{\itemsep}{0pt}%
                       \setlength{\parsep}{0pt}%
                      }%
                }%
               {\addtocounter{UnnumSchachtelebene}{-1}%
                \end{list}}
\makeatother
\newlength{\fktdefhilfslaenge}
\newcommand{\ohnefktdef}[4]
           {\hspace*{\fill}
            $\begin{array}[t]{ccc}%
            #1 & \nach & #2 \\
            #3 & \auf  & #4
            \end{array}$
            \hspace*{\fill}}
\newcommand{\fktdef}[5]
           {\hspace*{\fill}
            $\begin{array}[t]{cccc}%
            #1: & #2 & \nach & #3 \\    
                & #4 & \auf  & #5
            \end{array}$
            \settowidth{\fktdefhilfslaenge}{$#1$:}
            \hspace*{0.6 \fktdefhilfslaenge}  
            \hspace*{\fill}}
\newcommand{\fktdefpur}[5]
           {$\begin{array}[t]{cccc}%
            #1: & #2 & \nach & #3 \\    
                & #4 & \auf  & #5
            \end{array}$}
\newcommand{\fktdefabgesetztpur}[5]
           {
            
            $\begin{array}[t]{cccc}%
            #1: & #2 & \nach & #3 \\    
                & #4 & \auf  & #5
            \end{array}$
            \settowidth{\fktdefhilfslaenge}{$#1$:}
            \hspace*{0.6 \fktdefhilfslaenge}
            
           }
\newcommand{\fktdefabgesetzt}[5]
           {
           
            \hspace*{\fill}
            $\begin{array}[t]{cccc}%
            #1: & #2 & \nach & #3 \\    
                & #4 & \auf  & #5
            \end{array}$
            \settowidth{\fktdefhilfslaenge}{$#1$:}
            \hspace*{0.6 \fktdefhilfslaenge}  
            \hspace*{\fill}
            
            }
\newcommand{\ohnefktdefabgesetzt}[4]
           {      

            \hspace*{\fill}
            $\begin{array}[t]{ccc}%
            #1 & \nach & #2 \\
            #3 & \auf  & #4
            \end{array}$
            \hspace*{\fill}

            }
\newcommand{\doppelohnefktdefabgesetzt}[6]
           {

            \hspace*{\fill}
            $\begin{array}[t]{ccccc}%
            #1 & \nach & #2 & \nach & #3\\
            #4 & \auf  & #5 & \auf  & #6
            \end{array}$
            \hspace*{\fill}

            }
\newcommand{\anhang}%
           {\appendix
            \sectioninh{Anhang}
            \renewcommand{\Abschnittnummer}{%
                  \ifx\ZaehlerbisEbene\Ebenea{\Alph{section}}%
                  \else{%
                        \ifx\ZaehlerbisEbene\Ebeneb{\Alph{section}.\arabic{subsection}}%
                        \else{\Alph{section}.\arabic{subsection}.\arabic{subsubsection}}%
                        \fi}%
                  \fi}%
            \renewcommand{\Abschnittnummerpunkt}{\Abschnittnummer.}     
            }            
\newcommand{\anhangengl}%
           {\appendix
            \sectioninh{Appendix}
            \renewcommand{\Abschnittnummer}{%
                  \ifx\ZaehlerbisEbene\Ebenea{\Alph{section}}%
                  \else{%
                        \ifx\ZaehlerbisEbene\Ebeneb{\Alph{section}.\arabic{subsection}}%
                        \else{\Alph{section}.\arabic{subsection}.\arabic{subsubsection}}%
                        \fi}%
                  \fi}%
            \renewcommand{\Abschnittnummerpunkt}{\Abschnittnummer.}     
            }

\newcounter{wdhlstufe}
\newcommand{\sectioninh}[1]%
           {\section*{#1}%
            \addcontentsline{toc}{section}{#1}}
\newcommand{\bezeichnung}[3]%
           {\begin{Einrueckungpur}{\hbox to 6em{#1}\hbox to 2.4em{\hfill#2}}
            #3
            \end{Einrueckungpur}}

\newcommand{\doppelteinfach}{e}

\newcommand{\ifdoppelt}[1]{\ifthenelse{\equal{\doppelteinfach}{d}}{#1}{}}
\newcommand{\ifeinfach}[1]{\ifthenelse{\equal{\doppelteinfach}{e}}{#1}{}}

\newlength{\querfhilfsl}              %

\newlength{\hll}

\DefThmUmgeb{Theorem}{Theorem}{\ZaehlerbisEbene}
\DefThmUmgeb{Definition}{Definition}{\ZaehlerbisEbene}
\DefThmUmgeb{Satz}{Theorem}{\ZaehlerbisEbene}
\DefThmUmgeb{Proposition}{Theorem}{\ZaehlerbisEbene}
\DefThmUmgeb{Lemma}{Theorem}{\ZaehlerbisEbene}
\DefThmUmgeb{Folgerung}{Theorem}{\ZaehlerbisEbene}
\DefThmUmgeb{Corollary}{Theorem}{\ZaehlerbisEbene}
\DefThmUmgeb{Vorschrift}{Definition}{\ZaehlerbisEbene}
\DefThmUmgeb{Construction}{Definition}{\ZaehlerbisEbene}
\DefSThmUmgeb{FormSatz}{Theorem}{\ZaehlerbisEbene}{\glqq Satz\grqq} 
\DefThmUmgeb{Vermutung}{Theorem}{\ZaehlerbisEbene}
\DefThmUmgeb{Conjecture}{Theorem}{\ZaehlerbisEbene}
\DefThmUmgeb{Konvention}{Definition}{\ZaehlerbisEbene}
\DefThmUmgeb{Feststellung}{Theorem}{\ZaehlerbisEbene}
\DefUnterNumThmUmgeb{DefinitionZusatzNum}{Definition}{\ZaehlerbisEbene}{DefZN}{Definition}
\DefBspUmgeb{Beispiel}{Beispiel}{subsubsection}
\DefBspUmgeb{Example}{Example}{subsubsection}
\DefBspUmgeb{Frage}{Frage}{section}
\DefBspUmgeb{Question}{Question}{section}
\DefBspUmgeb{Aufgabe}{Aufgabe}{section}
\DefBspUmgeb{Ziel}{Ziel}{section}
\DefBemUmgeb{Bemerkung}
\DefBemUmgeb{Remark}
\DefSBemUmgeb{OffeneFrage}{Offene Frage}
\newcommand{\bdf}{\begin{Definition}}
\newcommand{\edf}{\end{Definition}}
\newcommand{\bvorsch}{\begin{Vorschrift}}
\newcommand{\evorsch}{\end{Vorschrift}}
\newcommand{\bconst}{\begin{Construction}}
\newcommand{\econst}{\end{Construction}}
\newcommand{\bthm}{\begin{Theorem}}
\newcommand{\ethm}{\end{Theorem}}
\newcommand{\bsatz}{\begin{Satz}}
\newcommand{\esatz}{\end{Satz}}
\newcommand{\bprop}{\begin{Proposition}}
\newcommand{\eprop}{\end{Proposition}}
\newcommand{\blem}{\begin{Lemma}}
\newcommand{\elem}{\end{Lemma}}
\newcommand{\bfolg}{\begin{Folgerung}}
\newcommand{\efolg}{\end{Folgerung}}
\newcommand{\bcorr}{\begin{Corollary}}
\newcommand{\ecorr}{\end{Corollary}}
\newcommand{\bfest}{\begin{Feststellung}}
\newcommand{\efest}{\end{Feststellung}}
\newcommand{\bbew}{\begin{Beweis}}
\newcommand{\ebew}{\end{Beweis}}
\newcommand{\bpf}{\begin{Proof}}
\newcommand{\epf}{\end{Proof}}
\newcommand{\bwnum}{\begin{WeiterNumerierteListe}}
\newcommand{\ewnum}{\end{WeiterNumerierteListe}}
\newcommand{\bdfzn}{\begin{DefinitionZusatzNum}}
\newcommand{\edfzn}{\end{DefinitionZusatzNum}}
\newcommand{\bbem}{\begin{Bemerkung}}
\newcommand{\ebem}{\end{Bemerkung}}
\newcommand{\brem}{\begin{Remark}}
\newcommand{\erem}{\end{Remark}}
\newcommand{\bnum}{\begin{NumerierteListe}}
\newcommand{\enum}{\end{NumerierteListe}}
\newcommand{\bunum}{\begin{UnnumerierteListe}}
\newcommand{\eunum}{\end{UnnumerierteListe}}
\newcommand{\bbsp}{\begin{Beispiel}}
\newcommand{\ebsp}{\end{Beispiel}}
\newcommand{\bex}{\begin{Example}}
\newcommand{\eex}{\end{Example}}
\newcommand{\bfrag}{\begin{Frage}}
\newcommand{\efrag}{\end{Frage}}
\newcommand{\bquest}{\begin{Question}}
\newcommand{\equest}{\end{Question}}
\newcommand{\baufg}{\begin{Aufgabe}}
\newcommand{\eaufg}{\end{Aufgabe}}
\newcommand{\bof}{\begin{OffeneFrage}}
\newcommand{\eof}{\end{OffeneFrage}}
\newcommand{\bverm}{\begin{Vermutung}}
\newcommand{\everm}{\end{Vermutung}}
\newcommand{\bconj}{\begin{Conjecture}}
\newcommand{\econj}{\end{Conjecture}}
\newcommand{\bkonv}{\begin{Konvention}}
\newcommand{\ekonv}{\end{Konvention}}
\newcommand{\bglklein}{\begin{StandardUnnumGleichungKlein}}
\newcommand{\eglklein}{\end{StandardUnnumGleichungKlein}}
\newcommand{\bgl}{\begin{StandardUnnumGleichung}}
\newcommand{\egl}{\end{StandardUnnumGleichung}}
\newcommand{\bglrtext}{\begin{XrelYZNumGleichung}}
\newcommand{\eglrtext}{\end{XrelYZNumGleichung}}
\newcommand{\zgl}{\EineZeileGleichung}

\newcommand{\zglklein}[1]{\zgl{\textstyle#1}}
\newcommand{\znumgl}{\EineNumZeileGleichung}

\newcommand{\berlgl}{\begin{StandardUnnumGleichung}}
\newcommand{\eerlgl}{\end{StandardUnnumGleichung}}
\newcommand{\beinrueck}{\begin{Einrueckungpur}} 
\newcommand{\eeinrueck}{\end{Einrueckungpur}}
\newcommand{\beinflist}{\begin{EinfachListe}} 
\newcommand{\eeinflist}{\end{EinfachListe}}
\newcommand{\beq}{\begin{equation}}
\newcommand{\eeq}{\end{equation}}
\newcommand{\bhin}{\begin{Hinrichtung}}
\newcommand{\ehin}{\end{Hinrichtung}}
\newcommand{\brueck}{\begin{Rueckrichtung}}
\newcommand{\erueck}{\end{Rueckrichtung}}
\newcommand{\bvl}{\begin{AutoLabelLaengenListe}{\ListNullAbstaende}}
\newcommand{\evl}{\end{AutoLabelLaengenListe}}
\newcommand{\df}[1]{{\bf #1}}
\newcommand{\zglnum}[2]{\znumgl{#1\label{#2}}}
\chardef\tempcat=\the\catcode`\@
\catcode`\@=11

\def\@gobble#1{}
\def\@testgrave{\`}
\def\@stressit{\futurelet\chartest\@stresschar }

\def\@stresschar#1{%
  \ifx #1y\def\result{\futurelet\chartest\@yligature}%
  \else \ifx #1Y\def\result{\futurelet\chartest\@Yligature}%
  \else \ifx\chartest\@testgrave \def\result{\accent"26 }%
  \else \def\result{\accent"26 #1}%
  \fi \fi \fi
  \result }

\def\@yligature{%
  \ifx a\chartest \def\result{\accent"26 \char"1F \@gobble}%
  \else \ifx u\chartest \def\result{\accent"26 \char"18 \@gobble}%
  \else \def\result{\accent"26 y}%
  \fi \fi
  \result }

\def\@Yligature{%
  \ifx a\chartest \def\result{\accent"26 \char"17 \@gobble}%
  \else \ifx A\chartest \def\result{\accent"26 \char"17 \@gobble}%
  \else \ifx u\chartest \def\result{\accent"26 \char"10 \@gobble}%
  \else \ifx U\chartest \def\result{\accent"26 \char"10 \@gobble}%
  \else \def\result{\accent"26 Y}%
  \fi \fi \fi \fi
  \result }

\def\!{\ifmmode \mskip-\thinmuskip \fi}

\def\cyracc{\chardef\i="10%
  \def\cydot{{\kern0pt}}%
  \def\cprime{\char"7E }\def\Cprime{\char"5E }%
  \def\cdprime{\char"7F }\def\Cdprime{\char"5F }%
  \def\dbar{dj}\def\Dbar{Dj}%
  \def\dz{\char"1E }\def\Dz{\char"16 }%
  \def\dzh{\char"0A }\def\Dzh{\char"02 }%
  \def\'##1{\if c##1\char"0F %
    \else \if C##1\char"07 %
    \else \accent"26 ##1\fi \fi }%
  \def\`##1{\if e##1\char"0B %
    \else \if E##1\char"03 %
    \else \errmessage{accent \string\` not defined in cyrillic}%
        ##1\fi \fi }%
  \def\=##1{\if e##1\char"0D %
    \else \if E##1\char"05 %
    \else \if \i##1\char"0C %
    \else \if I##1\char"04 %
    \else \errmessage{accent \string\= not defined in cyrillic}%
        ##1\fi \fi \fi \fi }%
  \def\u##1{\if \i##1\accent"24 i%
    \else \accent"24 ##1\fi }%
  \def\"##1{\if \i##1\accent"20 \char"3D %
    \else \if I##1\accent"20 \char"04 %
    \else \accent"20 ##1\fi \fi }%
  \def\!{\ifmmode \def\result{\mskip-\thinmuskip}%
    \else \def\result{\@stressit}\fi \result}}

\def\keep@cyracc{\let\cyr=\relax \let\i=\relax
        \let\ubar=\relax \let\cydot=\relax
        \let\cprime=\relax \let\Cprime=\relax
        \let\cdprime=\relax \let\Cdprime=\relax
        \let\dbar=\relax \let\Dbar=\relax
        \let\dz=\relax \let\Dz=\relax
        \let\dzh=\relax \let\Dzh=\relax
        \let\'=\relax \let\`=\relax \let\==\relax
        \let\u=\relax \let\"=\relax \let\!=\relax }

\DeclareFontEncoding{OT2}{\cyracc}{}

\catcode`\@=\tempcat

\newcommand{\bignorm}[1]{\bigl\lVert#1\bigr\rVert}
\newcommand{\Bignorm}[1]{\Bigl\lVert#1\Bigr\rVert}
\newcommand{\cxy}[1]{{\cal V}_{#1}}
\newcommand{\was}{{\cal V}}            
\newcommand{\redukt}{{\cal R}}
\newcommand{\reg}{{\mathrm{reg}}}
\newcommand{\mi}[1]{\bm{#1}}
\newcommand{\tpd}[1]{\mi{#1}}      
\newcommand{\krd}{\delta}          
\newcommand{\anderedarst}{\rho}          
\newcommand{\bigdirsum}{\bigoplus}

\newcommand{\sws}{{\cal H}}
\newcommand{\hilb}{{\cal H}}

\newcommand{\pot}[3]{[#1_{#2}]^{\bullet #3}}
\newcommand{\web}{w}

\renewcommand{\darst}{\varphi}

\newlength{\adressabstand}
\begin{document}
\title{Proof of a Conjecture by Lewandowski and Thiemann}
\author{Christian Fleischhack\thanks{e-mail: 
            {\tt chfl@mis.mpg.de}} \\   
        \\
        {\normalsize\em Max-Planck-Institut f\"ur Mathematik in den
                        Naturwissenschaften}\\[\adressabstand]
        {\normalsize\em Inselstra\ss e 22--26}\\[\adressabstand]
        {\normalsize\em 04103 Leipzig, Germany}
        \\[-25\adressabstand]      
        {\normalsize\em Center for Gravitational Physics and Geometry}\\[\adressabstand]
        {\normalsize\em 320 Osmond Lab}\\[\adressabstand]
        {\normalsize\em Penn State University}\\[\adressabstand]
        {\normalsize\em University Park, PA 16802}
        \\[-25\adressabstand]}      
\date{March 30, 2003}
\maketitle
\begin{abstract}
It is proven that for compact, connected and semisimple 
structure groups every degenerate labelled web is strongly degenerate.
This conjecture by Lewandowski and Thiemann implies that 
diffeomorphism invariant operators 
in the category of piecewise smooth immersive paths
preserve the decomposition
of the space of integrable functions
w.r.t.\ the degeneracy and symmetry of the underlying labelled webs. 
This property is necessary for lifting these operators to well-defined operators
on the space of diffeomorphism invariant states.
\end{abstract}

\section{Introduction}
One of the most striking features of general relativity is its
invariance w.r.t.\ diffeomorphisms of the underlying space-time manifold.
Its implementation into the Ashtekar formulation, however, 
is still not fully worked out. Here, one considers objects like
generalized connections that are defined using finite graphs in the underlying
space or space-time. For technical purposes, one assumed in the very beginning
that these graphs are formed by piecewise analytic paths only. Namely,
only in this case two finite graphs are always both contained in some
third, bigger graph being again finite. This restriction has the drawback
that only analyticity preserving diffeomorphisms can be implemented into
that framework. In order to guarantee the inclusion of all diffeomorphisms,
at least, piecewise smooth and immersive paths have to be considered as well.
For the first time, this has been done by Baez and Sawin \cite{d3} introducing
so-called webs. These are certain collections of paths that are independent
enough to ensure the well-definedness of the generalized Ashtekar-Lewandowski
measure $\mu_0$. Applications to quantum geometry have then been 
studied first by Lewandowski and Thiemann \cite{e46}. For this purpose, they 
determined the set of possible parallel transports along webs and
then discussed the diffeomorphism group averaging to 
generate diffeomorphism invariant states. Here it turned out that 
the extension of the spin-network formalism to the smooth-case spin-webs
leads to degeneracies. These appear if some paths in a web share some 
full segment and the tensor product of their carried group representations
includes the trivial representation. They impede the spin webs to form 
an orthonormal basis of $\mu_0$-integrable functions -- 
a striking contrast to the spin-networks in the analytic case.
Moreover, the diffeomorphism averaging is defined only on those cylindrical 
functions that arise from nondegenerate spin-webs (having additionally
finite symmetry group). To define now diffeomorphism invariant operators
on diffeomorphism invariant states, these operators 
have to preserve the corresponding decomposition of integrable
functions w.r.t.\ their degeneracy. In \cite{e46}, Lewandowski and Thiemann
showed that the images of non-degenerate spin-webs under such operators
are at least still orthogonal to so-called strongly degenerate spin-webs. 
Now, they argued that these strongly degenerate spin-webs should be nothing but 
degenerate spin-webs, implying that diffeomorphism invariant operators
respect the non-degeneracy of webs. In this article we are going to prove
this conjecture.

The paper is organized as follows: After some preliminaries
we recall the terms ``richness'' and ``splitting'' from \cite{paper13}. 
They will be used to encode the relative position of (parts of) webs: 
do they coincide, are they in a certain sense independent? 
Next we study the decomposition of consistently parametrized paths 
into hyphs and list some properties of webs.
In Section \ref{sect:opvalprod} we provide
the technical details of the proof of the Lewandowski-Thiemann conjecture 
that will then be given in the subsequent section. In the final section of this
paper we study the ``canonical'' example \cite{d17,e46} of a degenerate
web.

\section{Preliminaries}
Let us briefly recall the basic facts and notations we need from the framework 
of generalized connections. General expositions can be found 
in \cite{a48,a30,a28} for the analytic framework. The smooth case
is dealt with in \cite{d3,d17,e46}. The facts on hyphs and
the conventions are due to \cite{paper3,diss,paper2+4}.

Let $\LG$ be some arbitrary Lie group (being compact from
Section \ref{sect:opvalprod} on) and $M$ be some manifold. 
Let $\Pf$ denote the set of all (finite) paths in $M$, i.e.\ the set of all
piecewise smooth and immersive mappings from $[0,1]$ to $M$.%
\footnote{Sometimes, for simplicity, we will speak about paths 
restricted to certain subintervals of $[0,1]$. 
By means of some affine map from that interval to $[0,1]$
we may regard these restrictions naturally as paths again.}
The set $\Pf$ is a groupoid
(after imposing the standard equivalence relation, i.e., saying that
reparametrizations and insertions/deletions of retracings are irrelevant).
A hyph $\hyph$ is some finite collection 
$(\gamma_1,\ldots,\gamma_n)$ of edges (i.e.\ non-selfintersecting paths)
each having a ``free'' point.
This means, 
for at least one direction none of the segments of $\gamma_i$ 
starting in that point in this direction is a full segment
of some of the $\gamma_j$ with $j<i$.
Graphs and webs are special hyphs.
The subgroupoid generated
by the paths in a hyph $\hyph$ will be denoted by $\KG\hyph$.
Hyphs are ordered in the natural way. In particular, $\hyph' \leq \hyph''$ 
implies $\KG{\hyph'} \teilmenge \KG{\hyph''}$. 
The set $\Ab$ of generalized connections $\qa$ is now defined by
\zglklein{\Ab := \varprojlim_\hyph \Ab_\hyph \iso \Hom(\Pf,\LG),}\noindent
with $\Ab_\gc := \Hom(\Pf_\gc,\LG)$ 
given the topology induced by that of $\LG$
for all finite tuples $\gc$ of paths. 
For those $\gc$ we define the (always continuous) map
$\pi_\gc : \Ab \nach \LG^{\elanz\gc}$ by $\pi_\gc(\qa) := \qa(\gc)$.
Note, that $\pi_\gc$ is surjective, if $\gc$ is a hyph.
Finally, for compact $\LG$,
the Ashtekar-Lewandowski measure $\mu_0$ is the unique
regular Borel measure on $\Ab$ whose push-forward $(\pi_\hyph)_\ast\mu_0$ 
to $\Ab_\hyph \iso \LG^{\elanz\hyph}$ 
coincides with the Haar measure there for every hyph $\hyph$.

\section{Richness and Splittings}
\label{sect:richsplit}
Let $n\in\N_+$ be some positive integer. 
We recall the notions ``richness'' and ``splitting'' from
\cite{paper13}. Proofs not presented in this section are either given in
\cite{paper13} or are obvious.

\bdf
\label{def:V_n}
We define
\bunum
\item
$\cxy n$ to be the set of all 
$n$-tuples with entries equal to $0$ or $1$ only;
\item
$\LG_v := \{(g^{v_1},\ldots,g^{v_n}) \mid g\in\LG\} \teilmenge \LG^n$
for every $v \in \cxy n$;      
and
\item
$\LG_V := \LG_{v^1} \:\cdots\: \LG_{v^k}$
for every ordered\footnote{By an ordered subset of $X$ we mean an arbitrary
tuple of elements in $X$ where every element in $X$ occurs at most once as
a component of that tuple. However, we will use the standard terminology 
of sets if misunderstandings seem to be impossible.}
subset $V = \{v^1,\ldots,v^k\} \teilmenge \cxy n$.
\eunum
\edf
We have, e.g., $\LG_{(1,0,1,0)} = \{(g,e_\LG,g,e_\LG) \mid g\in\LG\}$.

\subsection{Richness}
\bdf
\label{def:rich}
An ordered subset $V\teilmenge \cxy n$ is called \df{rich} iff
\bnum{2}
\item
for all $1 \leq i,j \leq n$ with $i\neq j$ there is an element $v\in V$ with $v_i \neq v_j$ and
\item
for all $1 \leq i \leq n$ there is an element $w\in V$ with $w_i \neq 0$.
\enum
\edf
For instance, let $n=4$. 
Then $V := \{(1,1,0,0),\:(1,0,1,0),\:(0,1,0,1),\:(0,0,1,1)\}$ 
is rich, but $\{(1,1,0,1),\:(1,0,1,1),\:(0,1,1,0)\}$ is not because it
fails to fulfill the first condition for $i=1$ and $j=4$. 

Next we quote the main theorem on rich ordered subsets from \cite{paper13}.
Note that every connected compact semisimple Lie group equals its
commutator subgroup.
\bthm
\label{thm:rich->full}
Let $\LG$ be a connected compact semisimple Lie group and $n$ be
some positive integer. Then there is a positive integer $q(n)$ 
such that $\pot\LG V{q(n)} = \LG^n$ for any rich ordered subset $V$ of $\cxy n$.
\ethm
Here, $\pot\LG V q$ denotes the $q$-fold 
multiplication $\LG_V\cdots\LG_V$ of $\LG_V$. On the other hand, we
use $\LG^n$ as usual for the $n$-fold direct product 
$\LG \kreuz \cdots \kreuz \LG$ of $\LG$.
Note, moreover, that $q(n)$ in the theorem above
does not depend on the ordering or the number
of elements in $V$. Finally, we have 
$\LG^n = \pot\LG V{q(n)} \teilmenge \pot\LG V q \teilmenge \LG^n$
for all $q \geq q(n)$.

\subsection{Splittings}

\bdf
\label{def:splitting}
\bunum
\item
A subset $V \teilmenge \cxy n$ is called 
\df{$n$-splitting} 
iff
\bnum{2}
\item
$\sum_{v\in V} v = (1,\ldots,1)$ and
\item
$(0,\ldots,0) \nichtin V$.
\enum
\item
Let $V$ and $V'$ be $n$-splittings. $V'$ is called \df{refinement}
of $V$ (shortly: $V' \geq V$) iff every $v \in V$ can be written
as a sum of elements in $V'$.
\eunum
\edf
Directly from the definition we get
\blem
\bunum
\item
We have $V \leq V_{\max}$ for all $n$-splittings $V$, where 
$V_{\max}$ contains precisely the elements of $\cxy n$ having precisely
one component equal $1$.
\item
An $n$-splitting $V$ is rich iff $V = V_{\max}$.
\eunum
\elem

\bdf
\label{def:pi_V}
For all $n$-splittings $V$ we define
\fktdefabgesetzt{\pi_V}{\LG^n}{\LG^n,}{(g_1,\ldots,g_n)}%
                {(g_{s_V(1)},\ldots,g_{s_V(n)})}
where $s_V(i)$ is given by 
\zgl{s_V(i) := \min\{j\in[1,n] \mid 
                  \text{there is a $v \in V$ with $v_j = 1 = v_i$}\}.}
\edf

\blem
\label{lem:eig(pi_V)}
We have for all $n$-splittings $V$ and $V'$ with $V \leq V'$:
\bnum{3}
\item
$s_{V'} \circ s_V = s_V$,
\item
$\pi_{V'} \circ \pi_V = \pi_V$,
\item
$\pi_V$ is a $\ast$-homomorphism and
\item
$\pi_{V_{\max}}$ is the identity.
\enum
\elem
\bpf
\bnum{3}
\item
Let $1 \leq i \leq n$ be given. Choose $v\in V$ and $v'\in V'$, such that
$v'_{s_V(i)} = 1 = v_{s_V(i)}$. By definition, we have
$v'_{s_{V'}(s_V(i))} = 1 = v'_{s_V(i)}$.
Due to $V \leq V'$, this implies $v_{s_{V'}(s_V(i))} = 1 = v_{s_V(i)}$,
hence $v_{s_{V'}(s_V(i))} = 1 = v_i$ by definition of $s_V$.
Again, by the minimum requirement in the definition of $s_V$ we have 
$s_V(i) \leq s_{V'}(s_V(i))$. On the other hand, $s_{V'}$ is obviously
non-increasing, hence $s_V(i) = s_{V'}(s_V(i))$.
\item
Follows immediately from $s_{V'} \circ s_V = s_V$.
\item
Clear by the properties of $n$-splittings.
\item
Trivial.
\qed
\enum
\epf

\blem
\label{lem:prodordsplit}
For every $n$-splitting $V$ 
we have $\LG_V = \prod_{v\in V} \LG_v = \pi_V(\LG^n)$
independently of the ordering in $V$. Moreover, $\LG_V$ is 
a Lie subgroup of $\LG^n$.
\elem

\bdf
\label{def:splitV(s)}
Let $n\in\N_+$ be some positive integer, $S$ be some set and
$\vec s$ be some $n$-tuple of elements of $S$.
Then the \df{splitting $V(\vec s)$ for $\vec s$} is given by
\zgl{V(\vec s) := 
     \{v\in \cxy n \mid 
      \text{$v_i = 1 = v_j$ $\aequ$ $s_i = s_j$}\}
      \setminus\{(0,\ldots,0)\}.}
\edf
For example, the splitting for $\vec s = (s_1,s_2,s_3,s_2)$
is $V(\vec s) = \{(1,0,0,0),(0,1,0,1),(0,0,1,0)\}$.

\blem
For every $n$, $S$ and $\vec s$ as given in Definition \ref{def:splitV(s)},
$V(\vec s)$ is an $n$-splitting.
\elem
\section{Consistent Parametrization}
In this short section, consistently parametrized paths \cite{d3} are studied. 
These are paths whose parameters coincide if their images in the manifold $M$
coincide. We will prove that those paths can always be decomposed at finitely
many parameter values such that the subpaths generated this way are
graph-theoretically (hence \cite{paper3} measure-theoretically) 
independent, unless they are equal.

\bdf
\label{def:nice}
\label{def:consist}
Let $\gc=(\gamma_1,\ldots,\gamma_n)$ be some $n$-tuple of edges.
\bunum
\item
$\gc$ is called \df{nice} iff its 
\df{reduction} $\redukt(\gc) := \{\gamma_1,\ldots,\gamma_n\}$ is a hyph.%
\footnote{Observe that $\{\gamma_1,\ldots,\gamma_n\}$ denotes
the (if necessary, ordered) {\em set}\/ of all components 
of the {\em tuple}\/ $\gc$.}
\item
$\gc$ is called \df{consistently parametrized} iff for all $i,j = 1,\ldots,n$   
we have 
\zgl{\gamma_i(t') = \gamma_j(t'') \breitrel\impliz t' = t''.} 
\eunum
\edf
For example, we have for $\gc = (\gamma_1,\gamma_2,\gamma_3,\gamma_4)$
with $\gamma_2 = \gamma_4$
\bgl
\redukt(\gc) & = & \{\gamma_1,\gamma_2,\gamma_3\}, \\
      V(\gc) & = & \{(1,0,0,0),(0,1,0,1),(0,0,1,0)\}.
\egl

\bprop
\label{prop:decom(const_param_paths)}
Let $\gc$ be a consistently parametrized $n$-tuple of edges 
and $I \teilmenge [0,1]$ be some nontrivial interval (i.e.\ $I$ consists
of at least two points).

Then there is some $N \in \N_+$ and a sequence
\zgl{\min I = \tau_0 < \tau_1 < \ldots < \tau_N = \max I,}
such that $\bigcup_{i=1}^N \redukt(\gc\einschr{[\tau_{i-1},\tau_i]})$
is a disjoint union and a hyph, i.e., in particular, each 
$\gc\einschr{[\tau_{i-1},\tau_i]}$ is nice.
\eprop
\bpf
\bunum
\item
Let $\gc = (\gamma_1,\ldots,\gamma_n)$. Define for every $\tau\in I$ 
and every $i=1,\ldots,n$ the sets
\zgl{
 I_{\tau,+,j,k} := 
   \{\tau'\in[\tau,1] \mid 
       \gamma_j\einschr{[\tau,\tau']} = \gamma_k\einschr{[\tau,\tau']}\} \cap I
}
and
\zgl{
 I_{\tau,-,j,k} := 
   \{\tau'\in[0,\tau] \mid 
       \gamma_j\einschr{[\tau',\tau]} = \gamma_k\einschr{[\tau',\tau]}\} \cap I.
}
Observe first, that $I_{\tau,\pm,j,k}$ is always closed, since 
edges are continuous mappings from $[0,1]$ to $M$ and $\gc$
is consistently parametrized. Moreover, it is always connected
and contains $\tau$ unless it is empty.
Consequently,
\zglnum{
 I_{\tau,\pm} \breitrel{:=}
   \bigcap_{\substack{j,k=1,\ldots,n \\ \text{$j\neq k$ and $I_{\tau,\pm,j,k} \setminus \{\tau\} \neq \leeremenge$}}}
        I_{\tau,\pm,j,k}
}{eq:defItau}
is always a closed and connected, but possibly empty subset of $I$.
(The sets $I_{\tau,\pm}$ are assumed empty, if 
$I_{\tau,\pm,j,k}\setminus\{\tau\} = \leeremenge$ for all $j \neq k$.)
More precisely, we have two cases. Excluding the
exception $\tau = \max I$, we have:
\bunum
\item
If $I_{\tau,+}$ is non-empty,
then $I_{\tau,+}$ is a nontrivial interval (i.e.\ not a single point)
because the intersection in \eqref{eq:defItau} is finite.%
\footnote{A finite intersection of intervals containing
$\tau$ and some other point larger than $\tau$ is again such an interval.}
\item
If $I_{\tau,+}$ is empty, 
then again by that finiteness
we have $I_{\tau,\pm,j,k} \setminus \{\tau\} = \leeremenge$, 
hence $\gamma_j\einschr{[\tau,1]} \notBB \gamma_k\einschr{[\tau,1]}$
for all $j\neq k$.
\eunum
Similar results are true for $I_{\tau,-}$.
\item
Assume first that there is some $\tau\in I$ such that $I_{\tau,+}$ 
(for $\tau \neq \max I$) or $I_{\tau,-}$ (for $\tau \neq \min I$)
is empty.
Then we have in the first case
$\gamma_j\einschr{[\tau,\max I]} \notBB \gamma_k\einschr{[\tau,\max I]}$
for all $j\neq k$,
hence, $\gc\einschr I \ident \redukt(\gc\einschr I)$ is a hyph.
Defining $\tau_0 := \min I$ and $\tau_1 := \max I$, we get the assertion.
The second case is completely analogous.
\item
Assume now that there is no $\tau\in I$ such that $I_{\tau,+}$  
(for $\tau \neq \max I$) or $I_{\tau,-}$ (for $\tau \neq \min I$)
is empty. 
\bunum
\item
Construction of the sequence $(\tau_i)$ in $I$

Set $\tau_0 := \min I$. Then proceed successively, until
$\tau_j = \max I$ for some $j$:
\bnum{3}
\item
$\tau_{2i+1} := \max I_{\tau_{2i},+}$.
\item
$\tau_{2i+2} := \max \{\tau \in [\tau_{2i+1},\max I] \mid 
                         I_{\tau_{2i},+} \cap I_{\tau,-} \neq \leeremenge\}$.
\enum
\item
Well-definedness of the construction
\bnum{2}
\item
$\tau_{2i+1}$ exists, since $I_{\tau_{2i},+}$ is always a closed interval.
Moreover, $\tau_{2i+1} > \tau_{2i}$, since by assumption $\tau_{2i} \neq \max I$
and $I_{\tau_{2i},+}$ is nonempty, hence a nontrivial interval starting
at $\tau_{2i}$.
\item
$\tau_{2i+2}$ exists. In fact, since by construction 
$\min I \leq \tau_{2i} < \tau_{2i+1} < \max I$ and so neither 
$I_{\tau_{2i},+}$ nor $I_{\tau_{2i+1},-}$ are 
empty, we get
$\tau_{2i+1} \in I_{\tau_{2i},+} \cap I_{\tau_{2i+1},-}$.
Hence, the set $J$ which $\tau_{2i+2}$ is supposed to be the maximum of, 
is non-empty.
It remains the question whether $J$ has indeed a maximum. 
For this, set $\sigma := \sup J$
and assume $\sigma_l\konvhoch \sigma$ strictly increasing 
with non-empty $I_{\tau_{2i},+} \cap I_{\sigma_l,-}$ for all $l\in\N$.
Fix $j\neq k$. There are two cases:
\bunum
\item
Let there exist some $l'$ such that
$\gamma_j\einschr{[\tau_{2i+1},\sigma_l]} \neq \gamma_k\einschr{[\tau_{2i+1},\sigma_l]}$
for all $l \geq l'$. 

Then $I_{\sigma,-,j,k} \setminus \{\sigma\}$ is empty: 
Otherwise,
there would be some $l_0 \geq l'$ such that $\sigma_{l_0} \in I_{\sigma,-,j,k}$,
and then $\sigma_{l_0} \in I_{\sigma_{l_0 + 1},-,j,k} \obermenge
I_{\sigma_{l_0 + 1},-} \ni \tau_{2i+1}$ which would imply that 
$\gamma_j\einschr{[\tau_{2i+1},\sigma_{l_0}]} = 
         \gamma_k\einschr{[\tau_{2i+1},\sigma_{l_0}]}$.
Contradiction. 
\item
Let there exist no $l'$ such that
$\gamma_j\einschr{[\tau_{2i+1},\sigma_l]} \neq \gamma_k\einschr{[\tau_{2i+1},\sigma_l]}$
for all $l \geq l'$. 

Then there is an infinite subsequence $(\sigma_{l_q})$ of $(\sigma_l)$, 
such that we have
$\gamma_j\einschr{[\tau_{2i+1},\sigma_{l_q}]} =
       \gamma_k\einschr{[\tau_{2i+1},\sigma_{l_q}]}$ for all $q$.
Hence, 
$\gamma_j\einschr{[\tau_{2i+1},\sigma]} =
       \gamma_k\einschr{[\tau_{2i+1},\sigma]}$, i.e.
$\tau_{2i+1} \in I_{\sigma,-,j,k}$.
\eunum
Altogether, since $I_{\sigma,-} \neq \leeremenge$ by assumption,
we have $\tau_{2i+1} \in I_{\sigma,-}$, and so $\sigma\in J$,
since $\tau_{2i+1} \in I_{\tau_{2i},+}$. 
Obviously, $\tau_{2i+2} \geq \tau_{2i+1}$. 
\enum
\item
Stopping of the Construction

Suppose, there were no $N \in \N$ such that $\tau_N = \max I$.
Then $(\tau_i)_{i\in\N}$ is a strictly increasing sequence in $I$
having some limit $\tau\in I$ with $\tau_i < \tau$ for all $i$. 
Of course, $\tau > \min I$. 

Let $\tau' \in I_{\tau,-}$ with $\tau' < \tau$. (Remember 
that $I_{\tau,-}$ is nonempty.) Then 
there is some $i_0\in\N$ with $\tau' \leq \tau_{2i_0+1} < \tau$.
Consequently, $I_{\tau_{2i_0},+} \cap I_{\tau,-}$ contains $\tau_{2i_0+1}$.
This implies by the second step of the construction above, that
$\tau \leq \tau_{2i_0+2}$. This, however, is a contradiction to 
$\tau > \tau_i$ for all $i$.
\item
Final adjustment

Drop now all $\tau_{2i+2}$ from that sequence with $\tau_{2i+1} = \tau_{2i+2}$,
and denote the resulting finite subsequence again by $(\tau_0,\ldots,\tau_N)$.
\eunum
This sequence fulfills the requirements of the proposition:
\bnum{2}
\item
$\redukt(\gc\einschr{[\tau_{i-1},\tau_i]})$ is a hyph. 

Let first $i$ correspond to some ``originally'' odd $i$. 
Choose some path in $\redukt(\gc\einschr{[\tau_{i-1},\tau_i]})$,
say $\gamma_j\einschr{[\tau_{i-1},\tau_i]}$.
If $\gamma_j \einschr{[\tau_{i-1},\tau_i]} \BB \gamma_k \einschr{[\tau_{i-1},\tau_i]}$,
then there is some $\sigma \in (\tau_{i-1},\tau_i]$ with
$\gamma_j \einschr{[\tau_{i-1},\sigma]} = \gamma_k \einschr{[\tau_{i-1},\sigma]}$,
hence $[\tau_{i-1},\sigma] \teilmenge I_{\tau_{i-1},+,j,k}$.
By construction, we have  
$I_{\tau_{i-1},+,j,k} \obermenge I_{\tau_{i-1},+} = 
  [\tau_{i-1},\tau_{i}]$ and thus
$\gamma_j \einschr{[\tau_{i-1},\tau_i]} = \gamma_k \einschr{[\tau_{i-1},\tau_i]}$.
This means, they define the same element in   
$\redukt(\gc\einschr{[\tau_{i-1},\tau_i]})$. Therefore,
$\gamma_j \einschr{[\tau_{i-1},\tau_i]} \notBB \gamma_k \einschr{[\tau_{i-1},\tau_i]}$
for different elements. By the consistent parametrization,
$\redukt(\gc\einschr{[\tau_{i-1},\tau_i]})$ is a hyph.

The case of ``even'' $i$ goes analogously.
\item
$\bigcup_i \redukt(\gc\einschr{[\tau_{i-1},\tau_i]})$ is a hyph
and a disjoint union. 

The consistent parametrization of $\gc$ implies
that $\gamma_j\einschr{[\tau_{i-1},\tau_i]} \BB
     \gamma_{j'}\einschr{[\tau_{i'-1},\tau_{i'}]}$.
(or any other relation $\EB$, $\BE$ or $\EE$)
is possible for $i = i'$ only.
Together with the previous step we get the assertion.
\qed
\enum
\eunum
\epf

\section{Basic Facts about Webs}
Let us start with some definitions. Note that the definition of 
the $\gc$-type of a point is slightly different from that in \cite{d3}.
\bdf
\label{def:regular}
Let $\gc$ be some $n$-tuple of paths.
\bunum
\item
A point $x\in M$ is called \df{$\gc$-regular} iff $x$ is not 
an endpoint or nondifferentiable point of one of the paths in $\gc$ and 
there is a neighbourhood
of $x$ whose intersection with $\im \gc$ is an embedded interval. \cite{d3}
\item
$\tau\in[0,1]$ is called 
$\gc$-regular iff $\gamma(\tau)$ is $\gc$-regular for all $\gamma\in\gc$.
\eunum
\edf
\bdf
\label{def:type}
Let $\gc$ be some $n$-tuple of paths.
\bunum
\item
For every $x \in M$ we define 
the \df{$\gc$-type} $v(x)\in\cxy n$ of $x$ by
\zgl{v(x)_i := \begin{cases} 
                  1 & \text{if $x\in\im \gamma_i$} \\
                  0 & \text{if $x\nichtin\im \gamma_i$}
               \end{cases} .}
\item
For every consistently parametrized $\gc$ 
we define 
\zglklein{
V_\gc := \bigcup_{\text{$\tau\in[0,1]$, $\tau$ $\gc$-regular}} V(\gc(\tau)).
}
\eunum
\edf
For consistently parametrized $\gc$, obviously, $V(\gc(\tau))$ 
is the set of all $\gc$-types of points in $\gc(\tau)$.
Note, moreover, that in general the set $V_\gc$ of types in $\gc$ and the 
splitting $V(\gc)$ for $\gc$ do not coincide. For instance, we have
in the case of Figure \ref{fig:stdex} (see page \pageref{fig:stdex} with
$\gc := (\gamma_1,\gamma_2,\gamma_3,\gamma_4)$)
\bgl
V_\gc & = & \{(1,1,0,0),(1,0,1,0),(0,1,0,1),(0,0,1,1)\}, \\
V(\gc) & = & \{(1,0,0,0),(0,1,0,0),(0,0,1,0),(0,0,0,1)\}.
\egl
\noindent
In a certain sense, $V_\gc$ is finer. $V(\gc)$ only looks whether two
whole paths are equal or not. $V_\gc$ looks closer at the image points 
of $\gc$.

We now recall the definition of tassels and webs 
owing to Baez and Sawin \cite{d3,d17}. 
\bdf
\label{def:web}
\bunum
\item
A finite ordered set $T = \{c_1,\ldots, c_n\}$ 
of paths is called \df{tassel based on} 
$p\in\im T$ iff the following conditions are met:
\bnum{5}
\item 
$\im T$ lies in a contractible open subset of $M$.
\item 
$T$ can be consistently parametrized in such a way 
that $c_i(0) = p$ is the left endpoint of every path $c_i$.  
\item 
Two paths in $T$ that intersect at a point other than $p$
intersect at a point other than $p$ in every neighborhood of $p$.
\item 
For every neighbourhood $U$ of $p$, 
any $T$-type which occurs at some regular point in $\im T$ occurs at
some regular point in $U \cap \im T$.
\item 
No two paths in $T$ have the same image.
\enum
\item
A finite collection $\web = \web^1 \cup \dots \cup \web^k$ 
of tassels is called \df{web} iff for all $i \neq j$ the following
conditions are met:
\bnum{3}
\item 
Any path in the tassel $\web^i$ intersects any path in 
$\web^j$, if at all, only at their endpoints.
\item 
There is a neighborhood of each such intersection point 
whose intersection with $\im(\web^i \cup \web^j)$ 
is an embedded interval.
\item 
$\im \web^i$ does not contain the base of $\web^j$.
\enum
\eunum
\edf
Next, we list some important properties of webs that can 
be derived immediately from statements in \cite{d3}.
The proofs are given in \cite{paper13}.
\bprop
\label{prop:eig(webs)}
For every web $w$ the set $[0,1]_\reg$ of $w$-regular parameter values 
is open and dense in $[0,1]$.
Moreover, 
the function $V(w(\cdot)) : [0,1]_\reg \nach \cxy{\elanz w}$, assigning
to every $w$-regular $\tau$ its splitting, is locally constant.
\eprop
\blem
\label{lem:tassel_rich}
For every web $w$ the set $V_w$ of $w$-types occurring in $w$ is rich.
\elem
Let us define the set
$
 \was(w) := 
      \bigcap_{\tau\in(0,1]} \bigcup_{\sigma\in[0,\tau]_\reg} \{V(w(\sigma))\}
$ 
of all those splittings $V(w(\sigma))$ that appear in every neighbourhood
of $0$. Here, $I_\reg$ denotes the set of $w$-regular elements 
in an arbitrary interval $I\teilmenge[0,1]$.
\blem
\label{lem:types_in_V(w)}
Let $w$ be a web. 
Then for all $v\in V_w$ there is some $V\in\was(w)$ with $v\in V$.
In particular, $\was(w)$ is nonempty (if $w$ is nonempty).
\elem

\bcorr
\label{corr:rich(V_w)}
$\bigcup_{V\in\was(w)} V$ equals $V_w$ for every web $w$ and is rich.
\ecorr

\section{Operator-Valued Integrals}
\label{sect:opvalprod}
Let now $\LG$ be {\em compact}.
Let us fix some positive integer $n$ and 
some $n$-tuple $\vec\darst = (\darst_1,\ldots,\darst_n)$
of irreducible (unitary) representations of $\LG$ with 
$X_i$ being the representation space of $\darst_i$.
Set $\tpd\darst := \bigtensor_k \darst_k$ to be the tensor product
representation of $\LG^n$ on $X := \bigtensor_k X_k$
corresponding to $\vec\darst$.
Moreover, let $Y := \quer X \tensor X$.
$Y$ is now the representation space for the
$\LG^n$-representation
$\quer{\tpd\darst} \tensor \tpd\darst = 
   \bigtensor_k \quer{\darst_k} \tensor \bigtensor_k \darst_k$.
Finally, we equip $\End\:X$ and $\End\:Y$ with the standard operator norm.
\bdf
\label{def:Q_D}
For every continuous function $D : \LG^n \nach \End\:X$ we define
\bunum
\item
the (integrated and normalized Frobenius) norm%
\footnote{Note, that ${\norm\cdot}_F$ is, in general, not a matrix norm
due to the normalization.}
of $D$ by 
\zgl{{\norm D}_F^2 := \inv{\dim X} \int_{\LG^n} \tr (D^\ast D) \: \dd\mu_\Haar}
and
\item
the operator $Q_D \in \End\:Y$ by
\zgl{Q_D := \int_{\LG^n} \quer D \tensor D \:\: \dd\mu_\Haar.}
\eunum
\edf

\blem
\label{lem:Fnorm_unit}
Let $D,E : \LG^n \nach \End\:X$ be continuous functions.

If $E$ is unitary, then ${\norm E}_F = 1$ and
${\norm{E^\ast D E}}_F = {\norm D}_F $.
\elem
\bpf
Trivial.
\qed
\epf

\blem
\label{lem:Q_D=proj}
Let $D : \LG^n \nach \End\:X$ be a $\ast$-homomorphism.

Then $Q_D : Y \nach Y$ is an orthogonal projector.
\elem
\bpf
By $Q_D^\ast = Q_{D^\ast}^{\phantom\ast}$ and $D^\ast(\vec g) = D(\vec g{}^\ast)$, 
the homomorphy property of $D$ implies
\bgl[2ex]
Q_D^\ast Q_D^{\phantom\ast} 
 & = & \Bigl(\int_{\LG^n} (\quer D{}^\ast \tensor D{}^\ast) (\vec g) \:\dd\mu_\Haar \Bigr) \:
       \Bigl(\int_{\LG^n} (\quer D \tensor D) (\vec g') \:\dd\mu_\Haar \Bigr) \s
 & = & 
       \int_{\LG^n} \int_{\LG^n} (\quer D \tensor D) (\vec g{}^\ast \vec g') 
               \:\dd\mu_\Haar \:\dd\mu_\Haar  
\breitrel=
       \int_{\LG^n} (\quer D \tensor D) (\vec g) \:\dd\mu_\Haar \s
 & = & Q_D
\egl
using the translation invariance and normalization of the Haar measure.
Hence, we get
$Q_D^\ast = (Q_D^\ast Q_D^{\phantom\ast})^\ast 
  = Q_D^\ast Q_D^{\phantom\ast} = Q_D^{\phantom\ast}$
and $Q_D^{\phantom\ast} = Q_D^{\phantom\ast} Q_D^{\phantom\ast}$.
\qed
\epf

\subsection{Scalar-Product Projectors}
\bdf
\label{def:P_V}
We define for all $n$-splittings $V$
\zgl{P_V := Q_{\tpd\darst\circ\pi_V}
             =  \int_{\LG^n} 
                     (\quer{\tpd\darst} \tensor \tpd\darst) \circ \pi_V
                     \:\dd\mu_\Haar \in \End\:Y}
and set $P_0 := P_{V_{\max}}$.
\edf
\blem
\label{lem:eig(P_V)}
We have for all $n$-splittings $V$ and $V'$:
\bnum{4}
\item
$P_{V'} P_V = P_{V'} = P_V P_{V'}$ if $V \leq V'$.
\item
$P_V$ is an orthogonal projection on $Y$.
\item
$P_0 P_V = P_0 = P_V P_0$.
\item
$\norm{P_V} = 1$.
\item
$P_V Y = \{y \in Y \mid \text{$y$ is 
  $(\quer{\tpd\darst} \tensor \tpd\darst) 
      (\LG_V)$-invariant}\}$.
\enum
\elem
\bpf
\bnum{4}
\item
Using Lemma \ref{lem:eig(pi_V)}
and the homomorphy property of $\tpd\darst$, we have
\zgl{
\tpd\darst(\pi_V(\vec g)) \: \tpd\darst(\pi_{V'}(\vec g')) 
 =  \tpd\darst(\pi_{V'}(\pi_V(\vec g))) \: \tpd\darst(\pi_{V'}(\vec g'))  
 =  \tpd\darst(\pi_{V'}(\pi_V(\vec g) \: \vec g')).
}
Consequently, 
\bgl[2ex]
P_{V} P_{V'} 
 & = & \Bigl(\int_{\LG^n} (\quer{\tpd\darst} \tensor \tpd\darst) (\pi_V(\vec g))
                     \:\dd\mu_\Haar \Bigr) \:
       \Bigl(\int_{\LG^n} (\quer{\tpd\darst} \tensor \tpd\darst) (\pi_{V'}(\vec g'))
                     \:\dd\mu_\Haar \Bigr) \s
 & = & \int_{\LG^n} \int_{\LG^n}
          (\quer{\tpd\darst} \tensor \tpd\darst) 
	    (\pi_{V'}(\pi_V(\vec g) \: \vec g'))
          \:\dd\mu_\Haar \: \dd\mu_\Haar  \s
 & = & \int_{\LG^n} 
          (\quer{\tpd\darst} \tensor \tpd\darst) 
	    (\pi_{V'}(\vec g'))
          \:\dd\mu_\Haar  \s
 & = & P_{V'},
\egl
where we used in the third step that the Haar measure is normalized and 
invariant w.r.t.\ $\vec g' \auf \pi_V(\vec g)^{-1} \vec g'$.
$P_{V'} P_V = P_V$ follows precisely the same way.
\item
Follows from Lemma \ref{lem:Q_D=proj} since each $\darst_k$ is unitary
and $\pi_V$ is a $\ast$-homo\-mor\-phism.
\item
Follows from $V \leq V_{\max}$ for all $V$ and the statements above.
\item
Being a projection, $\norm{P_V} = 1$ unless $P_V$ is zero.
Since $P_V P_0 = P_0$ and $P_0 \neq 0$ (for an explicit computation
of its matrix elements see the proof of Lemma \ref{lem:proj_sandwich}), 
$P_V = 0$ is impossible.
\item
Let $\phi_V : Y \nach \bigdirsum_l W_l$ be a unitary map decomposing
the $\LG^n$-representation
  $(\quer{\tpd\darst} \tensor \tpd\darst) \circ \pi_V$ 
into a direct sum of irreducible representations $\anderedarst_l$ on $W_l$.
Then we have 
\zglklein{
     \phi_V (P_V y)  \\
  \breitrel= (\phi_V \circ P_V \circ \phi_V^{-1})(\phi_V(y)) \\
  \breitrel=  \bigl(\int_{\LG^n} \bigdirsum_l \anderedarst_l \: \dd\mu_\Haar \bigr)
                  (\phi_V(y)). \\
}
Since $\int_{\LG^n} \anderedarst_l \: \dd\mu_\Haar$ equals $0$ if
$\anderedarst_l$ is non-trivial and equals $1$ if $\anderedarst_l$ is 
trivial, we have 
\bglklein
 &       & P_V y = y \\
 & \aequ & \phi_V (P_V y) = \phi_V (y) \\
 & \aequ & \phi_V(y) \text{ is contained in } 
              \bigdirsum_{\anderedarst_l = 0} \: W_l \\
 & \aequ & \phi_V(y) \text{ is invariant w.r.t.\ } 
              \bigdirsum_l \anderedarst_l (\LG^n) \\
 & \aequ & y \text{ is invariant w.r.t.\ }  
             \bigl((\quer{\tpd\darst} \tensor \tpd\darst) \circ \pi_V \bigr)(\LG^n)
              =
              \phi_V^{-1}
              \bigl(\bigdirsum_l \anderedarst_l (\LG^n)\bigr)
              \phi_V.
\eglklein
$\pi_V (\LG^n) = \LG_V$ gives the assertion.
\qed
\enum
\epf

\blem
\label{lem:proj_intersect}
Let $\was \teilmenge \cxy n$ be some subset and 
define $V_{\was} := \bigcup_{V\in\was} V$. Assume, moreover,
that there is some $q\in\N_+$ with $\pot\LG{V_{\was}}q  = \LG^n$.
Then we have $\bigcap_{V\in\was} P_V Y = P_0 Y$.
\elem
\bpf
\begin{SuperSet}
Since $P_V P_0 = P_0$, we have $P_0 Y \teilmenge P_V Y$ for all 
$V\in\cxy n\obermenge \was$. 
\end{SuperSet}
\begin{SubSet}
Let now $y\in P_V Y$ for all $V\in\was$.
By Lemma \ref{lem:eig(P_V)}, $y$ is invariant under each corresponding
$(\quer{\tpd\darst} \tensor \tpd\darst) (\LG_V)$,
hence w.r.t.\ $(\quer{\tpd\darst} \tensor \tpd\darst) (\LG_v)$ 
for all $v\in V_\was$.
By assumption, 
every element in $\LG^n$ can be written as some finite product of
elements in $\LG_{V_\was}$, hence in $\bigcup_{v\in V_\was} \LG_v$ as well.
By the homomorphy property of
$\tpd\darst$, we get the invariance of $y$ w.r.t.\
$(\quer{\tpd\darst} \tensor \tpd\darst) (\LG^n)$,
hence $y\in P_0 Y$.
\qed
\end{SubSet}
\epf

\subsection{More General Operators}
\blem
\label{lem:proj_sandwich}
For every continuous $D : \LG^n \nach \End\:X$ we have
$P_0 Q_D P_0 = {\norm D}_F^2 \: P_0$. 
\elem
\bpf
Introducing some bases on the $X_i$ and then forming multi-indices
we have 
\bglklein[1ex]
(P_0)^{\mi i\mi m}_{\mi j\mi n}
 & = & \int_{\LG^n} (\quer{\tpd\darst} \tensor \tpd\darst)^{\mi i\mi m}_{\mi j\mi n}
                     \: \dd\mu_\Haar
\breitrel= 
       \prod_k \int_\LG (\quer{\darst_k})^{i_k}_{j_k} 
                    (\darst_k)^{m_k}_{n_k} \: \dd\mu_\Haar 
\s & = & 
       \prod_k \inv{\dim \darst_k}\: \krd^{i_k m_k} \krd_{j_k n_k}
\breitrel= 
       \inv{\dim X} \: \krd^{\mi i\mi m} \krd_{\mi j\mi n}   
\eglklein
and hence 
\bglklein[1ex]
(P_0 Q_D P_0)^{\mi i\mi m}_{\mi j\mi n}
   & = & (P_0)^{\mi i\mi m}_{\mi p\mi r} \:
         (Q_D)^{\mi p\mi r}_{\mi q\mi s} \:
         (P_0)^{\mi q\mi s}_{\mi j\mi n} 
   \breitrel= 
         \inv{\dim X} \: \krd^{\mi i\mi m} \krd_{\mi p\mi r} \:
         (Q_D)^{\mi p\mi r}_{\mi q\mi s} \:
         \inv{\dim X} \: \krd^{\mi q\mi s} \krd_{\mi j\mi n}
\s & = & 
         \inv{\dim X} \: \krd^{\mi i\mi m} \krd_{\mi j\mi n} \: \:
         \inv{\dim X} \: (Q_D)^{\mi p\mi p}_{\mi q\mi q} 
   \breitrel=
         (P_0)^{\mi i\mi m}_{\mi j\mi n} \:
         \inv{\dim X} \: (Q_D)^{\mi p\mi p}_{\mi q\mi q}. 
\eglklein
Using
\zglklein{
(Q_D)^{\mi p\mi p}_{\mi q\mi q} 
  = \int_{\LG^n} \quer{D^{\mi p}_{\mi q}} D^{\mi p}_{\mi q} \:\: \dd\mu_\Haar
  = \int_{\LG^n} (D^\ast)^{\mi q}_{\mi p} D^{\mi p}_{\mi q} \:\: \dd\mu_\Haar
  = (\dim X) \: {\norm D}_F^2,
}   
we have 
$P_0 Q_D P_0 = {\norm D}_F^2 \: P_0$.
\qed
\epf

\bdf
\label{def:D_(V,k)}
Let $V$ be some $n$-splitting
and let $\bigdirsum_{l_k} \anderedarst_{k,l_k}$ 
for each $k=1,\ldots,n$ be the decomposition of 
$\bigtensor_{i : s_V(i) = k} \darst_i$ into irreducible $\LG$-representations.
Furthermore, denote the representation space of each $\anderedarst_{k,l_k}$ by
$W_{k,l_k}$, and let 
$\phi : X \nach \bigtensor_{k=s_V(k)} \bigdirsum_{l_k} W_{k,l_k}$ be
the corresponding unitary intertwiner. 

Define $D_{V,q} : \LG^n \nach \End\:X$  for all $1 \leq q \leq n$ via
\zgl{
\phi \: D_{V,q} (g_1,\ldots,g_n) \: \phi^{-1}
   := \bigtensor_{\substack{k \\ k = s_V(k)}} 
           \begin{cases}
             \bigdirsum_{l_k\phantom{\neq 0}} \anderedarst_{k,l_k} (g_k) & \text{ for $s_V(k) \neq s_V(q)$} \\
             \bigdirsum_{l_k\neq 0} \anderedarst_{k,l_k} (g_k) & \text{ for $s_V(k)  =   s_V(q)$} 
           \end{cases}
}
and set $Q_{V,q} := Q_{D_{V,q}}$.
\edf
In other words, $D_{V,q}$ just projects to the subspace of $X$ which is
orthogonal to the subspace that carries the trivial representation
after tensoring all $\darst_i$ where $i$ is ``equivalent'' to $q$,
i.e.\ where $i$ is running over all components 
in $v$ being $1$ 
where $v$ is just the element in $V$ whose $q$-component is $1$.
Note, furthermore, that $D_{V,q} = D_{V,q}  \circ \pi_V$.

\blem
\label{lem:Fnorm_degen}
For every $n$-splitting $V$ and every $1 \leq q \leq n$ we have
\zgl{{\norm{D_{V,q}}}_F^2 = 1 - \frac{d^0_{q}}{d_{q}}}
where $d_{q}$ is the dimension of the representation 
$\bigdirsum_{l} \anderedarst_{s_V(q),l}$ and $d_{q}^0$ the 
number of trivial $\anderedarst_{s_V(q),l}$ in this direct sum.
\elem
\bpf
Since $D_{V,q} = D_{V,s_V(q)}$, we may assume $q = s_V(q)$.
Then, using the unitarity of $\phi$ and $\darst_k$ and the fact that tensor products for
terms depending on different $g_k$ contribute to the norm as 
separate factors, we have 
\zgl{
 {\norm{D_{V,q}}}_F^2 = 
   \frac{\int_\LG \tr 
           \bigl(\bigdirsum_{l \text{ with } \anderedarst_{q,l} \neq 0} 
             \anderedarst_{q,l}^\ast \anderedarst_{q,l}^{\phantom\ast}\bigl) \: \dd\mu_\Haar}
        {\prod_{i : s_V(i) = q} \dim \darst_i}
 =  \frac{\sum_{l \text{ with } \anderedarst_{q,l} \neq 0} \dim \anderedarst_{q,l}}
        {\sum_l \dim \anderedarst_{q,l}}
 =  1 - \frac{d^0_{q}}{d_{q}}.
}
\qed
\epf

\blem
For every $n$-splitting $V$ and every $1 \leq q \leq n$ we have
$\norm{Q_{V,q}} \leq 1$.
\elem
\bpf
By construction, $D_{V,q}$ is a $\ast$-homomorphism. 
Now, Lemma \ref{lem:Q_D=proj} gives the assertion.
\qed
\epf

\subsection{Application to Nice Sets of Paths}
\blem
\label{lem:int_nice}
For every nice $n$-tuple $\gc$ of edges and every continuous 
$f : \LG^n \nach \C$ we have
\zgl{\int_\Ab f \circ \pi_\gc \:\dd\mu_0
    = \int_{\LG^n} f \circ \pi_{V(\gc)} \:\dd\mu_\Haar^n.}
\elem

\bpf
Assume $\gc$ nice and, w.l.o.g., 
$\redukt(\gc) = \{\gamma_1,\ldots,\gamma_k\}$. Then
\bgl
\pi_\gc(\qa) 
  & = & \bigl(h_\qa(\gamma_1),\ldots,h_\qa(\gamma_k),
              h_\qa(\gamma_{k+1}),\ldots,h_\qa(\gamma_n)\bigr) \\
  & = & \pi_{V(\gc)}
        \bigl(h_\qa(\gamma_1),\ldots,h_\qa(\gamma_k),
              h_\qa(\gamma_{k+1}),\ldots,h_\qa(\gamma_n)\bigr) \\
  & = & \pi_{V(\gc)}
        \bigl(h_\qa(\gamma_1),\ldots,h_\qa(\gamma_k),e_\LG,\ldots,e_\LG\bigr) \\
  & = & \pi_{V(\gc)} (\pi_{\redukt(\gc)} \kreuz 1_{n-k})
          (\qa).
\egl
Since, by assumption, $\redukt(\gc)$ is a hyph and $\mu_\Haar$ is normalized, 
we get the assertion from $(\pi_{\redukt(\gc)})_\ast \mu_0 = \mu_\Haar^k$.
Since the Haar measure is permutation invariant, we get the proof for
arbitrary $\redukt(\gc)$.
\qed
\epf

\bcorr
\label{corr:int_proj_nice}
For every nice $n$-tuple $\gc=(\gamma_1,\ldots,\gamma_n)$ of edges 
we have
\zgl{P_{V(\gc)} = \int_\Ab (\quer{\tpd\darst} \tensor \tpd\darst) 
                     \circ \pi_\gc \:\dd\mu_0.}
\ecorr

\section{Conjecture of Lewandowski and Thiemann}
First we recall very briefly the definition of spin webs and then
the two different notions of degeneracy \cite{e46}. 
The conjecture of Lewandowski and Thiemann will say that both are equivalent.
Throughout the whole section, let $\LG$ be compact.
\bdf
\label{def:spin_web}
\bunum
\item
A \df{spin web} $(w,\vec\darst)$ consists of a web $w$ and some
$\elanz w$-tuple $\vec\darst$ of (equivalence classes of) irreducible 
representations of $\LG$.
\item
The \df{spin web state} $(T_{w,\vec\darst})^{\mi i}_{\mi j}$ 
to a spin web $(w,\vec\darst)$ is defined
by 
\zgl{(T_{w,\vec\darst})^{\mi i}_{\mi j} := 
          \tpd\darst^{\mi i}_{\mi j} \circ \pi_w : \Ab \nach \C}
with the tensor-matrix functions
\fktdefabgesetzt{\tpd\darst^{\mi i}_{ \mi j} = %
                   \bigtensor_k (\darst_k)^{i_k}_{j_k}}
                {\LG^{\elanz \vec\darst}}{\C.}%
                {\vec g}{\prod_k \darst_k(g_k)^{i_k}_{j_k}}
\item
The \df{spin web space} $\sws_{w,\vec\darst}$ for the spin web $(w,\vec\darst)$
is the $\C$-linear span of all spin web states for $(w,\vec\darst)$.
The web space $\sws_w$ is defined to be the closure of the $\C$-span 
of all possible spin web states to the web $w$.
\eunum
\edf
We remark that the definition above can be extended directly from webs to hyphs.

Before we come to the definition of degeneracy, we still have to define
for every edge $s$ the projection $p_s : \hilb \nach \hilb$ 
\label{def:p_s}
as follows \cite{e46}: Let first $e$ be an edge and $\Psi \in \sws_e$. 
Then, $p_s \Psi := \Psi$ if $e$ and $s$ are disjoint (maybe up to their endpoints), 
and $p_s \Psi := (T_{e,0},\Psi) T_{e,0} \ident (1,\Psi) 1$ if $e$ is a
nontrivial subpath of $s$. 
This means, $p_s$ projects onto the part in $\hilb_s$ carrying the
trivial representation.
For the general case, let $\hyph = \{\gamma_1,\ldots,\gamma_n\}$ 
be some hyph with $\hyph \geq \{s\}$ and let 
$\Psi = \bigtensor \Psi_k \in \bigtensor \sws_{\gamma_k}$,
then $p_s \Psi := \bigtensor p_s \Psi_k$. One immediately checks
that $p_s$ is well defined. Thus, we may extend this definition
by linearity and continuity. 

\bdf
A splitting $V$ is called \df{$\vec\darst$-degenerate} iff
there is some $v\in V$ such that the decomposition of
$\bigtensor_{k : v_k = 1} \darst_k$ into irreducible $\LG$-representations
contains the trivial representation.
\edf
For example, let $\LG = SU(2)$ whose irreducible representations
are labelled by half-integers. Then $\{(1,1,0,0),(0,0,1,1)\}$ is
$(\inv2,\inv2,3,\frac52)$-degenerate, since $\inv2\tensor\inv2 \iso 1 \oplus 0$.

\bdf
\label{def:degen}
\bunum
\item
A spin web $(w,\vec\darst)$ is called \df{(weakly) degenerate}
iff there is some $w$-regular $\tau\in[0,1]$ such that 
$V(w(\tau))$ is $\vec\darst$-degenerate, i.e.\
there is some $w$-regular point $x\in\im w$ such that 
the trivial representation is contained in the decomposition of 
$\bigtensor_{j : x\in\im w_j} \darst_j$ into irreducible
representations.
\item
A spin web $(w,\vec\darst)$ is called \df{strongly degenerate}
iff there is a sequence $(s_l)_{l\in\N}$ of disjoint 
$w$-regular segments in $w$ such that 
\zgl{\lim_{l\gegen\infty} (1 - p_{s_0}) \cdots (1 - p_{s_l}) \Psi = 0}
for all $\Psi\in \sws_{w,\vec\darst}$. 
\eunum
\edf
Here, a $w$-regular segment equals $w_q\einschr I$ for some $w_q\in w$ 
and some interval $I\teilmenge[0,1]_\reg$.

Let us now state 
\bthm[Lewandowski-Thiemann Conjecture]
\label{thm:LTconj}
Let $\LG$ be compact, connected and semisimple.

Then a spin web is weakly degenerate iff it is strongly degenerate.
\ethm

\blem
\label{lem:durchschnitt(proj)}
Let $\LG$ be compact, connected and semisimple.

Then we have $\bigcap_{V\in\was(w)} P_V Y = P_0 Y$
for every web $w$.
\elem
\bpf
Since $V_w = \bigcup_{V\in\was(w)} V$ is rich by 
Corollary \ref{corr:rich(V_w)}, and since
$\LG$ is compact, connected and semisimple, 
Theorem \ref{thm:rich->full} guarantees that
$\pot\LG{V_w}q = \LG^n$ for some $q\in\N_+$. 
Now, Lemma \ref{lem:proj_intersect} gives the proof.
\qed
\epf

\bpf[Theorem \ref{thm:LTconj}]
Let first $(w,\vec\darst)$ be some spin web that is not weakly degenerate.
Then $p_s \Psi = 0$ for all $\Psi \in \sws_{w,\vec\darst}$ and all
$w$-regular segments $s$ in $w$. 
Consequently, 
\zgl{\lim_{l\gegen\infty} (1 - p_{s_0}) \cdots (1 - p_{s_l}) \Psi = \Psi,}
whence $(w,\vec\darst)$ is not strongly degenerate.

Let now $(w,\vec\darst)$ be some weakly degenerate spin web. 
Since the proof of its strong degeneracy 
is much more technical, we proceed in several steps.
\bnum{6}
\item
Notations

We denote the elements of 
$\was(w)$ by $V_1,\ldots,V_N$. Since $(w,\vec\darst)$ is weakly degenerate,
there is some $v \in V_w$, such that $\bigtensor_{k : v_k = 1} \darst_k$
contains the trivial representation. By Lemma \ref{lem:types_in_V(w)},
there is some $W\in\was(w)$ with $v\in W$. Finally,
let $1\leq q\leq n$ be some number with $v_q = 1$,
where $n$ as usual is the number of paths in $w$.
\item
Decomposition of $w$

Let us construct a sequence $(\tau_i)$ in $[0,1]$ that will be used
for the decomposition of $w$. 
For this, we first define inductively a strictly 
decreasing sequence $(\sigma_{i,j})_{i\in\N,0\leq j\leq N}$
as follows ($\sigma_{-1,N} := 1$):
\bnum{2}
\item
$\sigma_{i+1,0}$ is some $w$-regular element in $[0,\sigma_{i,N})$,
such that $V(w(\sigma_{i+1,0})) = W$;
\item
$\sigma_{i,j+1}$ is some $w$-regular element in $[0,\sigma_{i,j})$,
such that $V(w(\sigma_{i,j+1})) = V_{j+1}$.
\enum
By construction,
such $\sigma_{i,j}$ always exist and $\sigma_{i,j} > 0$ for all $i,j$.

Since $\sigma_{i,j}$ is always regular 
and the splitting function $[0,1]_\reg\ni\tau \auf V(w(\tau))$ is locally
constant, 
there are regular $\sigma^\pm_{i,j}$ such that 
\bunum
\item
the splitting function on $[\sigma^-_{i,j},\sigma^+_{i,j}]\ni\sigma_{i,j}$
is constant (i.e.\ equal to $V(w(\sigma_{i,j}))$);
\item
$\sigma^+_{i,0} > \sigma^-_{i,0} > \sigma^+_{i,1} > \sigma^-_{i,1}
 > \sigma^+_{i,2} > \ldots > \sigma^-_{i,N-1} > \sigma^+_{i,N}
 > \sigma^-_{i,N} > \sigma^+_{i+1,0}$ for all $i$.
\eunum
Now we decompose, according to Proposition \ref{prop:decom(const_param_paths)}, 
those intervals in $[0,1]$ that remain after
removing all the intervals $[\sigma^-_{i,j},\sigma^+_{i,j}]$.
More precisely, there are $N_{i,j} \in \N_+$ and $\tau_{i,j,k} \in [0,1]$
for $i,j,k\in\N$ with $0\leq j \leq N$ and $0\leq k \leq N_{i,j}$, such that
for all $i,j$
\bunum
\item
$\sigma^-_{i,j-1} = \tau_{i,j,0} > \tau_{i,j,1} > \ldots 
  > \tau_{i,j,N_{i,j}} = \sigma^+_{i,j}$;
\item
$\redukt(w\einschr{[\tau_{i,j,k+1},\tau_{i,j,k}]})$ is a hyph
for $k=0,\ldots,N_{i,j}-1$.
\eunum
Here, we have been quite sloppy with the notation in the case that $i$ or $j$
are getting out of range. In these cases, we extended our definitions naturally,
i.e., $\sigma_{i,-1}^- := \sigma_{i-1,N}^-$ and $\sigma_{0,-1} := 1$.

To simplify the notation we denote the members of the sequence
\zgl{(\tau_{0,0,0},\:\tau_{0,0,1},\:\ldots,\:\tau_{0,0,N_{0,0}},\:
          \tau_{0,1,0},\:\ldots,\:\ldots,\:\tau_{0,N,N_{0,N}},\:\tau_{1,0,0},\:\ldots)}
by $(\tau_0, \tau_1, \tau_2, \ldots)$. Additionally, we 
define $a_i\in\N$ for every $i$ by $\tau_{a_i} = \tau_{i,0,N_{i,0}}$.
This is precisely the endpoint of the $(2i+1)$-st%
\footnote{Note that, since $W$ is also 
a member of the sequence $V_1,\ldots,V_N$, it has two tasks and 
occurs therefore roughly twice as often as the other $V_i$s.
In fact, first it will be used to pick up the degeneracy and
second it will be used to make the sequence $V_1,\ldots,V_N$ rich.
Hence, we will need $W$ partially in the terms below that are affected by $p_s$
and partially in those that are not.}
interval (i.e., $[\tau_{a_i+1},\tau_{a_i}]$)
in our construction having splitting $W$.
Finally, we define 
\zgl{
I_{i} \breitrel{:=} [\tau_{i+1},\tau_{i}] \breitrel{\text{ and }}
J_{i} \breitrel{:=} [0,\tau_{i}],
}
and set 
\zgl{
V(i) \breitrel{:=} V(w\einschr{I_{i}}).
}
\item
Properties of the decomposition

We have for all $i,i'\in\N$:
\bnum{2}
\item
$\redukt(w\einschr{I_{i}})$ is a hyph.

If $I_i$ corresponds to some interval $[\tau_{s,j,k+1},\tau_{s,j,k}]$
with $k \neq N_{s,j}$, this follows directly from the construction.
Otherwise, i.e.\ for $I_i = [\sigma^-_{s,j},\sigma^+_{s,j}]$,
the assertion follows because $I_i$ then contains $w$-regular 
elements only and the splitting function is constant on $I_i$.
Therefore, by the consistent parametrization, the paths in 
$w\einschr{I_i}$ are disjoint or equal, proving the hyph property.
\item
$\redukt(w\einschr{J_{i}}) = w\einschr{J_{i}}$ is a web, 
hence a hyph as well.

To see this, use that 
$\dach w\einschr{[0,\tau]}$ is a web again 
for all webs $\dach w$ and all $\tau>0$.
\item
$w\einschr{I_{i}} \cap w\einschr{I_{i'}} \neq \leeremenge$
iff $i = i'$.

This is a consequence of the consistent parametrization of $w$.
\item
$w\einschr{I_{i}} \cap w\einschr{J_{i'}} = \leeremenge$
for $i<i'$.

This comes from the consistent parametrization again.
\item
Performing the multiplication with decreasing indices, we have
\zglklein{
 w 
 \breitrel=  w\einschr{J_{i+1}} \: 
               \prod_{i'=i}^0 w\einschr{I_{i'}} 
 \breitrel\ident  w\einschr{J_{i+1}} \circ
               w\einschr{I_{i}} \circ \cdots \circ w\einschr{I_{0}}
}
directly from the definitions above.
\item
$\redukt(w\einschr{J_{i+1}}) \cup
 \bigcup_{i'=0}^i \redukt(w\einschr{I_{i'}})$ is a hyph.

Since each reduction involved is a hyph itself, this comes from the 
consistent parametrization.
\enum

\item
Estimation of products of projections

Let $\varepsilon$ be given. 
Consider the set $\was := \bigcup_{i} \{V(i)\}$
of all splittings occurring in the above decomposition.
Of course, $\was$ is finite,
because there are only finitely many $n$-splittings at all.
Moreover, $\was(w) \teilmenge \was$, and every $V_l \in \was(w)$
occurs infinitely often in $(I_{0}, I_{1}, \ldots)$.
Since every $P_V$ is a projection (Lemma \ref{lem:eig(P_V)})
and since $\bigcap_{V\in\was(w)} P_V Y = P_0 Y$ (Lemma \ref{lem:durchschnitt(proj)}),
Proposition \ref{prop:prod(proj)} guarantees that
for every $i\in\N$ there is some integer $K(i,\varepsilon) > i$, such that 
\zglklein{
 \bignorm{\prod_{i'=K(i,\varepsilon)}^{i+1} P_{V(i')} - P_0} < \varepsilon.
} 
Since $P_V P_0 = P_0 = P_0 P_V$ and $\norm{P_V} = 1$ for all $V$,
we get 
\zglklein{
 \norm{P_{V(i_-)} \cdots P_{V(i_+)} - P_0} < \varepsilon
} 
for all $i_\pm$ with
$i_- \geq K(i,\varepsilon) \geq i+1 \geq i_+$.
Choose now a strictly increasing sequence 
$(l_0^\varepsilon,l_1^\varepsilon,\ldots)$ in $\N$
fulfilling 
\zgl{a_{l_{\nu+1}^\varepsilon} > K(a_{l_\nu^\varepsilon},\varepsilon_\nu)
 \:\: \text{ with } \:\: \varepsilon_\nu := (1+\varepsilon)^{1/2^{\nu+2}}-1}
for all $\nu\in\N$.
For starting, we set $l_{-1}^\varepsilon := -1$ and $a_{-1} := -1$.

Since $K(l,\cdot) > l$ for all $l$, we have 
$a_{l_{\nu+1}^\varepsilon} > a_{l_\nu^\varepsilon}$,
i.e.\ indeed a strictly increasing sequence $(l_\nu^\varepsilon)$.
Moreover, we have 
$a_{l_{\nu+1}^\varepsilon} - 1 \geq K(a_{l_\nu^\varepsilon},\varepsilon_\nu) 
\geq a_{l_\nu^\varepsilon} + 1$.
Consequently, by $\norm{P_0} = 1$, $\norm{Q_{W,q}} \leq 1$ and 
Proposition \ref{prop:absch_altprod}, 
we have for all $L\in\N$

\bglklein
\bignorm{P_0 \prod_{\nu=L}^0\bigl(Q_{W,q}
       P_{V(a_{l_{\nu}^\varepsilon}-1)} \cdots P_{V(a_{l_{\nu-1}^\varepsilon}+1)} 
       \bigr)  -
      P_0 \prod_{\nu=L}^0\bigl(Q_{W,q} P_0 \bigr) } & < & \varepsilon.
\eglklein
Let us consider the second 
product. By Lemma \ref{lem:proj_sandwich} 
we have 
\bgl
\norm{P_0 (Q_{W,q} P_0 )^{L+1}}
 & = & \norm{P_0 ( Q_{W,q} P_0 P_0 )^{L+1}}
 \breitrel= \norm{(P_0 Q_{W,q} P_0)^{L+1} \: P_0 } \\
 & \leq & \norm{P_0 Q_{W,q} P_0}^{L+1}
 \breitrel= \norm{D_{W,q}}_F^{2(L+1)}.
\egl
Due to the choice of $W$ and $q$, we have 
$\norm{D_{W,q}}_F < 1$ by Lemma \ref{lem:Fnorm_degen}.
Thus, there is some $L(\varepsilon)\in\N$, such that 
\bgl
\bignorm{P_0 \: (Q_{W,q}P_0)^{L(\varepsilon)+1}} & < & \varepsilon.
\egl
Consequently,
\zglnum{
\Bignorm{P_0 \:\:
       \prod_{\nu=L(\varepsilon)}^0\Bigl( Q_{W,q}
             P_{V(a_{l_{\nu}^\varepsilon}-1)} \cdots P_{V(a_{l_{\nu-1}^\varepsilon}+1)}
       \Bigr)} 
 \breitrel<  2 \varepsilon.
}{eq:absch1}
\item
Application to the spin web $(w,\vec\darst)$

We have for all $i' \in \N$
\zglklein{
T_{w,\vec\darst} 
  = \tpd\darst \circ \pi_w 
  = \bigl(\tpd\darst \circ \pi_{w\einschr{J_{i'+1}}}\bigr) \:\cdot\:
    \prod_{i = i'}^0 \tpd\darst \circ \pi_{w\einschr{I_i}}.
}
Set now $s_i := w_q\einschr{I_{a_i}}$,
i.e., $s_i$ is the 
restriction of $w_q$ to $[\sigma^-_{i,0},\sigma^+_{i,0}]$
which is just the $(2i+1)$-st interval in our originally chosen sequence whose 
corresponding splitting is $W$.
Extending the action of $p_s$ naturally from the spin web states
$(T_{w,\vec\darst})^{\mi i}_{\mi j}$
to the corresponding operators $T_{w,\vec\darst}$, we get
\bgl[2ex]
 &  & (1 - p_{s_{l_0}}) \cdots (1 - p_{s_{l_L}}) T_{w,\vec\darst} \\
 &= & \bigl(\tpd\darst \circ \pi_{w\einschr{J_{a_{l_L}+1}}}\bigr) \:\cdot\:
       \prod_{i = a_{l_L}}^0 
        (1 - p_{s_{l_0}}) \cdots (1 - p_{s_{l_L}})
        (\tpd\darst \circ \pi_{w\einschr{I_i}}) \s
 &= & \bigl(\tpd\darst \circ \pi_{w\einschr{J_{a_{l_L}+1}}}\bigr) \:\cdot\:
       \prod_{\nu=L}^0 
         \Bigl(
         (1 - p_{s_{l_\nu}}) (\tpd\darst \circ \pi_{w\einschr{I_{a_{l_\nu}}}}) \: \cdot 
           \prod_{i = a_{l_\nu}-1}^{a_{l_{\nu-1}}+1} 
                (\tpd\darst \circ \pi_{w\einschr{I_i}}) \Bigr) 
\egl
for all strictly increasing (finite) sequences $(l_0,\ldots,l_L)$, where
w.l.o.g.\ $l_{-1} = -1$. Thus, we get
\bgl
&   & \int_\Ab \quer{(1 - p_{s_{l_0}}) \cdots (1 - p_{s_{l_L}}) T_{w,\vec\darst}}
               \tensor          
               (1 - p_{s_{l_0}}) \cdots (1 - p_{s_{l_L}}) T_{w,\vec\darst} 
               \:\:\dd\mu_0     \\
& = & \int_\Ab \quer{(\tpd\darst \circ \pi_{w\einschr{J_{a_{l_L}+1}}})}
               \tensor
               (\tpd\darst \circ \pi_{w\einschr{J_{a_{l_L}+1}}}) 
               \:\:\dd\mu_0    \:\:\cdot  \\
&   & \:\:\cdot\:\: 
      \prod_{\nu=L}^0 \Bigl(
         \int_\Ab \quer{(1-p_{s_{l_\nu}})(\tpd\darst \circ \pi_{w\einschr{I_{a_{l_\nu}}}})}
                  \tensor
                  {(1-p_{s_{l_\nu}})(\tpd\darst \circ \pi_{w\einschr{I_{a_{l_\nu}}}})}
                  \:\:\dd\mu_0    \:\:\cdot  \\
&   & \:\:
      \cdot \: \:
         \prod_{i = a_{l_\nu}-1}^{a_{l_{\nu-1}}+1} 
         \int_\Ab \quer{(\tpd\darst \circ \pi_{w\einschr{I_i}})} 
                  \tensor 
                 (\tpd\darst \circ \pi_{w\einschr{I_i}}) 
                  \:\:\dd\mu_0  \Bigr)  
      \erl{Corollary \ref{corr:prod_int_op}} \\
& = & P_0 
      \:\cdot\:
      \prod_{\nu=L}^0 \Bigl(Q_{W,q} \cdot 
         \prod_{i = a_{l_\nu}-1}^{a_{l_{\nu-1}}+1} P_{V(i)} \Bigr).
      \erl{Lemma \ref{lem:int_nice} and 
           Corollary \ref{corr:int_proj_nice}} 
\egl
Here we used that $V(w\einschr{I_i}) = V(i)$ 
and $V(a_{l_\nu}) = W$. Moreover, we exploited the definitions
of $Q_{W,q}$ (Definition \ref{def:D_(V,k)}) and 
$p_s$ (page \pageref{def:p_s}) to replace the $(1-p_{s_l})$-terms by
$Q_{W,q}$.
Finally, note
that $w\einschr{J_i}$ is always a web, hence
$V(w\einschr{J_i}) = V_0$.  

Let now $(T_{w,\vec\darst})^{\mi i}_{\mi j}$ be some spin web state
for $(w,\vec\darst)$. 
Then 
\bgl[1ex]
 &   & \norm{(1 - p_{s_{l_0}}) \cdots (1 - p_{s_{l_L}}) (T_{w,\vec\darst})^{\mi i}_{\mi j}}^2 \s
 & = & \skalprod{(1 - p_{s_{l_0}}) \cdots (1 - p_{s_{l_L}}) (T_{w,\vec\darst})^{\mi i}_{\mi j}}%
                {(1 - p_{s_{l_0}}) \cdots (1 - p_{s_{l_L}}) (T_{w,\vec\darst})^{\mi i}_{\mi j}} \\
 & = & \Bigl(
       \int_\Ab \quer{(1 - p_{s_{l_0}}) \cdots (1 - p_{s_{l_L}}) T_{w,\vec\darst}}
                \tensor          
                (1 - p_{s_{l_0}}) \cdots (1 - p_{s_{l_L}}) T_{w,\vec\darst} 
                \:\:\dd\mu_0
       \Bigr)^{\mi i\mi i}_{\mi j\mi j}  
\egl
is just some matrix element of the above operator on $Y$.
Since $Y$ is a finite-dimensional Hilbert space, all norms are equivalent, hence
there is some constant $C\in\R$ (depending only on $Y$ and the norms fixed 
from the beginning), such that 
\bgl
 &   & 
\norm{(1 - p_{s_{l_0}}) \cdots (1 - p_{s_{l_L}}) (T_{w,\vec\darst})^{\mi i}_{\mi j}}^2 
\\
 & & \hspace*{8em} \breitrel\leq 
    C \:\: \Bignorm{P_0 
      \:\cdot\:
      \prod_{\nu=L}^0 \Bigl(Q_{W,q} \cdot 
         \prod_{i = a_{l_\nu}-1}^{a_{l_{\nu-1}}+1} P_{V(i)} \Bigr)} \:.
\egl
\item
Final step: Proof of the Lewandowski-Thiemann conjecture

Let $\varepsilon > 0$ be given. 
Choose $(l_0^\varepsilon,l_1^\varepsilon,\ldots)$ as above.
Then there is some $L(\varepsilon)$, such that \eqref{eq:absch1}
is fulfilled. 
Consequently, setting $N(\varepsilon) := l_{L(\varepsilon)}^\varepsilon$
we have
\bgl
\norm{(1 - p_{s_{0}}) \cdots 
      (1 - p_{s_{N(\varepsilon)}}) 
      (T_{w,\vec\darst})^{\mi i}_{\mi j}}^2  
& \leq & \norm{(1 - p_{s_{l_0^\varepsilon}}) \cdots 
         (1 - p_{s_{l_{L(\varepsilon)}^\varepsilon}}) 
         (T_{w,\vec\darst})^{\mi i}_{\mi j}}^2 \\
& < & 2C\varepsilon
\egl
because $(1-p_s)$ is a projection. Moreover, we used that 
$(1-p_{s'})$ and $(1-p_{s''})$ commute, if $\im s'$ and $\im s''$ are 
disjoint. Note that $C$ does not depend on $\varepsilon$, but only on the
fixed spin web. 

Hence, 
$\lim_{l\gegen\infty} 
       (1 - p_{s_{0}}) \cdots 
       (1 - p_{s_{l}}) 
       (T_{w,\vec\darst})^{\mi i}_{\mi j}  = 0$ 
for all $\mi i, \mi j$. By linearity we get
\bgl 
\lim_{l\gegen\infty} (1 - p_{s_{0}}) \cdots (1 - p_{s_{l}}) \Psi & = & 0
\egl
for all $\Psi\in\sws_{w,\vec\darst}$.
\qed
\enum
\epf
We remark finally that the Lewandowski-Thiemann conjecture can be extended 
even to arbitrary connected compact Lie groups $\LG$ -- with one restriction,
of course: In general, it is only true for webs $w$ where $V_w$ generates
full $\R^{\elanz w}$. In fact, then we have $\pot\LG{V_w}q = \LG^n$ 
for some $q\in\N$. \cite{paper13}
This has been the crucial ingredient for the proof 
of Lemma \ref{lem:durchschnitt(proj)}.
In the proof of the
Lewandowski-Thiemann conjecture itself, the assumption of semisimplicity 
has been used
only indirectly to guarantee the applicability of the lemma just mentioned.

\section{``Standard'' Example of a Web}
The original idea \cite{e46} of Lewandowski and Thiemann to prove 
their conjecture was that it should always be possible to find
degenerate segments $s_l$, such that -- in our terminology -- 
the portion of the web between two subsequent intervals corresponds
always to $P_0$, which is given if these portions are 
measure-theoretically, i.e.\ 
in a certain sense ``strongly'' independent.
They argued that, for that purpose, it ought to be sufficient to prove just
the holonomical independence of these portions. Unfortunately,
this is not the case as we will see in this section. Therefore, 
the article \cite{paper13}, 
where the holonomical independence has been established, 
cannot prove the Lewandowski-Thiemann conjecture yet.
However, all this is not a real problem, since we have now been able to prove 
in the
present article that these portions can be chosen, such that the corresponding
operators are sufficiently close to $P_0$ which still gives the proof.

In this final section we consider $\LG = SU(2)$.
Let now $V_1 := \{(1,1,0,0),(0,0,1,1)\}$ and
$V_2 := \{(1,0,1,0),(0,1,0,1)\}$ be two $4$-splittings. Moreover,
let the quadrupel $\vec\darst = (\inv2,\inv2,\inv2,\inv2)$
consist of spin-$\inv2$ representations of $SU(2)$.

\blem
\label{lem:Pprod-dblbbl}
We have $P_{V_1} Y \cap P_{V_2} Y = P_0 Y$, but $P_{V_1} P_{V_2} \neq P_0$.
\elem
\bpf
Of course, $V := V_1 \cup V_2$ is rich. Hence, by Theorem \ref{thm:rich->full},
we have $\pot\LG V {q(4)} = \LG^4$. Lemma \ref{lem:proj_intersect} now gives
$P_{V_1} Y \cap P_{V_2} Y = P_0 Y$.

Let us now prove $P_{V_1} P_{V_2} \neq P_0$.
We have for every integrable function $f$ on $\LG^4$
\bgl[2ex]
&   & \int_{\LG^{4}} f \bigl(
         g_{1+}g_{2+}, \: g_{1+}g_{2-}, \: g_{1-}g_{2+}, \: g_{1-}g_{2-}
         \bigr) \: \dd\mu_\Haar^{4} \s
& = & \int_{\LG^{4}} f \bigl(
         g_{1+}g_{1-}g_{2+}, \: g_{1+}g_{1-}g_{2-}, \:
         g_{1-}g_{2+}, \: g_{1-}g_{2-}
         \bigr) \: \dd\mu_\Haar^{4} \\
&   & \erl{Translation invariance w.r.t. $g_{1+} \auf g_{1+}g_{1-}$} \\
& = & \int_{\LG^{3}} f \bigl(
         g_{1+}g_{2+}, \: g_{1+}g_{2-}, \: g_{2+}, \: g_{2-}
         \bigr) \: \dd\mu_\Haar^{3} \\
&  &         \erl{Translation invariance w.r.t. $g_{2\pm} \auf (g_{1-})^{-1}g_{2\pm}$; Normalization} \\
& = & \int_{\LG^{3}} f \bigl(
         g_{1}g_{2}, \: g_{1}g_{3}, \: g_{2}, \: g_{3}
         \bigr) \: \dd\mu_\Haar^{3}.       \erl{Renumeration}
\egl
Consequently,
\bgl[2ex]
&   & \bigl(P_{V_1} P_{V_2}\bigr)^{\mi i \mi k}_{\mi j \mi l} \\
& = & \int_{\LG^4} 
        \quer{(g_{1+} g_{2+})^{i_1}_{j_1} \: (g_{1+} g_{2-})^{i_2}_{j_2} \:
              (g_{1-} g_{2+})^{i_3}_{j_3} \: (g_{1-} g_{2-})^{i_4}_{j_4}} \:\:\:
\cdot \\[-1ex] &   & \phantom{\int_{\LG^{11}} \varphi \bigl(\varphi \bigl(} \:\:\:\: \cdot \:\:
        (g_{1+} g_{2+})^{k_1}_{l_1} \: (g_{1+} g_{2-})^{k_2}_{l_2} \:
        (g_{1-} g_{2+})^{k_3}_{l_3} \: (g_{1-} g_{2-})^{k_4}_{l_4} \:
      \:\: \dd\mu_\Haar^4 \s
& = & \int_{\LG^3}
        \quer{(g_1 g_2)^{i_1}_{j_1} \:
              (g_1 g_3)^{i_2}_{j_2} \: (g_2)^{i_3}_{j_3} \: 
              (g_3)^{i_4}_{j_4}} \:\: 
        (g_1 g_2)^{k_1}_{l_1} \:
        (g_1 g_3)^{k_2}_{l_2} \: (g_2)^{k_3}_{l_3} \: 
        (g_3)^{l_4}_{l_4} \:
      \:\: \dd\mu_\Haar^3 \s
& = & \int_{\LG^3}
        \quer{(g_1)^{i_1}_{m_1}} \:   \quer{(g_2)^{m_1}_{j_1}} \:\:\:
        \quer{(g_1)^{i_2}_{m_2}} \:   \quer{(g_3)^{m_2}_{j_2}} \:\:\:
        \quer{(g_2)^{i_3}_{j_3}}\:\:\:\quer{(g_3)^{i_4}_{j_4}} \:\:
\cdot \\[-1ex] &   & \phantom{\int_{\LG^{11}} \varphi \bigl(\varphi \bigl(} \:\:\:\: \cdot \:\:
             {(g_1)^{k_1}_{n_1}} \:        {(g_2)^{n_1}_{l_1}} \:\:\:
             {(g_1)^{k_2}_{n_2}} \:        {(g_3)^{n_2}_{l_2}} \:\:\:
             {(g_2)^{k_3}_{l_3}}\:\:\:     {(g_3)^{k_4}_{l_4}} \:\:
      \:\: \dd\mu_\Haar^3 \\
& = & \skalprod{g{}^{i_1}_{m_1} g{}^{i_2}_{m_2}}%
               {g{}^{k_1}_{n_1} g{}^{k_2}_{n_2}}_{\Haar,1} \:\:
      \skalprod{g{}^{m_1}_{j_1} g{}^{i_3}_{j_3}}%
               {g{}^{n_1}_{l_1} g{}^{k_3}_{l_3}}_{\Haar,2} \:\:
      \skalprod{g{}^{i_4}_{j_4} g{}^{m_2}_{j_2}}%
               {g{}^{k_4}_{l_4} g{}^{n_2}_{l_2}}_{\Haar,3}. \:\:
\egl
Now, we set $j_4 := l_2 := 2$ and the remaining indices 
of $P_{V_1} P_{V_2}$ equal to $1$:
\bglklein[1ex]
&   & \bigl(P_{V_1} P_{V_2}\bigr)^{1111\:1111}_{1112\:1211} \s
& = & \skalprod{g{}^{1}_{m_1} g{}^{1}_{m_2}}%
               {g{}^{1}_{n_1} g{}^{1}_{n_2}}_{\Haar,1} \:\:
      \skalprod{g{}^{m_1}_{1} g{}^{1}_{1}}%
               {g{}^{n_1}_{1} g{}^{1}_{1}}_{\Haar,2} \:\:
      \skalprod{g{}^{1}_{2} g{}^{m_2}_{1}}%
               {g{}^{1}_{1} g{}^{n_2}_{2}}_{\Haar,3} \s
& = & \sum_{m_1,m_2} 
      \skalprod{g{}^{1}_{m_1} g{}^{1}_{m_2}}%
               {g{}^{1}_{m_1} g{}^{1}_{m_2}}_{\Haar,1} \:\:
      \skalprod{g{}^{m_1}_{1} g{}^{1}_{1}}%
               {g{}^{m_1}_{1} g{}^{1}_{1}}_{\Haar,2} \:\:
      \skalprod{g{}^{1}_{2} g{}^{m_2}_{1}}%
               {g{}^{1}_{1} g{}^{m_2}_{2}}_{\Haar,3} \s
& = & \sum_{m_1,m_2} \inv6 (1 + \krd_{m_1 m_2}) \: \inv6 (3-m_1) \: \inv6 (-1)^{m_2+1} \s
& = & \inv{6^3}. \\
\eglklein
Here, in the second step we used that, by Lemma \ref{lem:skalp1}, 
only those scalar products are non-zero,
where the sum of the first two upper (lower) indices equals that
of the last two upper (lower) indices. Thus, by the third scalar product, 
only $m_2 = n_2$ contributes. Analogously, $m_1 = n_1$ by
the second scalar product. Finally, we used Lemma \ref{lem:skalp2} and
Lemma \ref{lem:skalp3}.
Since, as seen in the proof of Lemma \ref{lem:proj_sandwich}, we have 
$(P_0)^{1111\:1111}_{1112\:1211} = 0$, we get 
$P_{V_1} P_{V_2} \neq P_0$.
\qed
\epf
Before stating the final result of this paper, let us recall
\bdf
Let $\gc$ be some tuple of paths.
\bunum
\item
$\gc$ is called \df{measure-theoretically independent} iff
$\pi_\gc{}_\ast \mu_0 = \mu_\Haar^{\elanz\gc}$.
\item
$\gc$ is called \df{holonomically independent} iff 
for every $\vec g\in\LG^{\elanz\gc}$ there is some smooth connection $A\in\A$
such that $h_A(\gc) = \vec g$.
\eunum
\edf
Note that the holonomical independence of $\gc$ is independent of the ultralocal
trivialization chosen to define the group values $h_A(\gc)$
of parallel transports for $A$.
\begin{figure}\begin{center}
\epsfig{figure=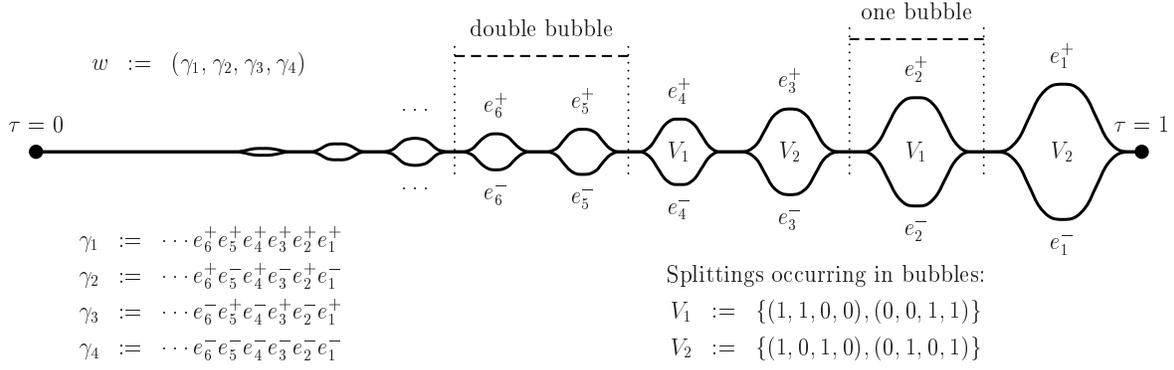,scale=0.75}
\caption{Standard Example of a Web \cite{d17,e46}}
\label{fig:stdex}
\end{center}\end{figure}
\bprop
Let $\LG = SU(2)$ and let $w$ be the web of Figure \ref{fig:stdex},
where each of the four paths in $w$ is labelled by the $\inv2$-representation
of $SU(2)$. Then we have:
\bnum{2}
\item
This spin web $(w,\vec\darst)$ is weakly degenerate.
\item
$w\einschr I$ is holonomically independent, but
{\em not}\/ measure-theoretically independent
for every interval $I\teilmenge(0,1]$ whose image under $w$
contains at least four subsequent bubbles.
\enum
\eprop
We remark that
$w\einschr I$ is measure-theoretically independent if and only if
$0$ is contained in $I$ (and $I$ is nontrivial, of course). 
\bpf
\bunum
\item
The weak degeneracy of $(w,\vec\darst)$ is clear.
\item
$w\einschr I$ is not measure-theoretically independent.

Applying the terminology of Section \ref{sect:opvalprod} to the case
of the given spin web, we see that 
\zgl{
\skalprod{\tpd\darst{}^{\mi i}_{\mi j}\circ\pi_{w\einschr I}}{\tpd\darst{}^{\mi k}_{\mi l}\circ\pi_{w\einschr I}}
= \bigl(P' (P_{V_1} P_{V_2})^{B} P''\bigr)^{\mi i \mi k}_{\mi j \mi l}.
}

Here, $V_1$ and $V_2$ are again given as above. These are precisely the
two splittings that occur in $w$ for $w$-regular parameter values.
$P'$ is the identity, if the bubble,
that is (at least partially, but nontrivially) passed first by $w\einschr I$ 
(when running through $I$ with increasing parameter values),
corresponds to splitting $V_1$. It equals $P_{V_2}$ otherwise. Analogously,
$P''$ is the identity, if the last (partially) passed bubble is of splitting
$V_2$, and equals $P_{V_1}$ otherwise. Finally, $B$ is the number of
double bubbles of ``type'' $(V_1,V_2)$
passed by $w\einschr I$ (one bubble may be passed only 
partially). Note that $I$ does not contain $0$, hence $B$ is indeed finite.

If $w\einschr I$ were measure-theoretically independent, we would
get 
\zgl{\skalprod{\tpd\darst{}^{\mi i}_{\mi j}\circ\pi_{w\einschr I}}{\tpd\darst{}^{\mi k}_{\mi l}\circ\pi_{w\einschr I}}
 = \skalprod{\tpd\darst{}^{\mi i}_{\mi j}}{\tpd\darst{}^{\mi k}_{\mi l}}_{\LG^4}
 = (P_0)^{\mi i \mi k}_{\mi j \mi l}.}
This, however, is a contradiction since, 
by Lemma \ref{lem:Pprod-dblbbl}, we know that $P_{V_1} P_{V_2} \neq P_0$,
hence $P' (P_{V_1} P_{V_2})^{B} P'' \neq P_0$ by Lemma \ref{lem:two_proj}.
\item
$w\einschr I$ is holonomically independent.

As one checks quite easily, 
we have 
$\LG_{V_1} \LG_{V_2} \LG_{V_1} \LG_{V_2} = \LG^4 = 
   \LG_{V_2} \LG_{V_1} \LG_{V_2} \LG_{V_1}$
for every connected semisimple $\LG$.
Consequently, the results shown in
\cite{paper13} imply that if two double bubbles (i.e.\ twice the sequence
$(V_1,V_2)$ or $(V_2,V_1)$ of splittings) are passed, then the web,
restricted to these two double bubbles, is strongly holonomically
independent. Since $w\einschr I$ passes at least two double bubbles,
we get the assertion.
\qed
\eunum
\epf

\section{Acknowledgements}
The author thanks Jerzy Lewandowski for fruitful discussions.
The author has been supported by the Reimar-L\"ust-Stipendium
of the Max-Planck-Gesellschaft and in part by NSF grant PHY-0090091.

\anhangengl
\enlargethispage{0.1cm}
\section{Convergence of Projector Products}
\bprop
\label{prop:prod(proj)}
Let $H$ be a finite-dimensional Hilbert space 
and let $P_1, \ldots, P_n$ be (self-adjoint) projections on $H$.
Moreover, let $H_i := P_i H$, $i = 1,\ldots,n$, be the 
corresponding projection spaces. 
Now, define $H_0 := \bigcap_{i=1}^n H_i$ and denote the 
projector from $H$ to $H_0$ by $P_0$.
Next, let $I \teilmenge \{1,\ldots,n\}$ be some subset,
such that $\bigcap_{i\in I} H_i = H_0$.
Finally, let $(j_k)_{k\in\N_+}$ be a sequence of integers, such that
\bunum
\item
$1 \leq j_k \leq n$ for all $k \in \N_+$;
\item
every $i\in I$ occurs infinitely many times in $(j_k)_{k\in\N}$.
\eunum
Then both $\prod_{k=1}^N P_{j_k}$ and $\prod_{k=N}^1 P_{j_k}$ converge
for $N \gegen \infty$ in the operator norm to $P_0$.
\eprop
\bpf
First let us assume $H_0 = 0$.
\bunum
\item
Let a nonempty subset $L \teilmenge \{1,\ldots,n\}$ be called {\em full}\/ 
iff $\bigcap_{i\in L} H_i = 0$. Then by \cite{m1}
for all full $L$
there is some constant $\vartheta_L \in [0,1)$, such that
\zgl{\norm{P_{l_1} P_{l_2} \cdots P_{l_N}} \leq \vartheta_L}
for all $N$ and 
for all finite sequences $l_1,\ldots,l_N$ of elements in $L$ where every
element of $L$ occurs at least once.%
\footnote{If $\elanz L = 1$, i.e. $L = \{i\}$,
then $H_i = 0$ and $P_i =0$. Consequently,
$\norm{P_{l_1} P_{l_2} \cdots P_{l_N}} = \norm{P_i^N} = 0 =: \vartheta_L < 1$
for all sequences $l_1,\ldots,l_N$.}
\item
The number of full subsets $L \teilmenge \{1,\ldots,n\}$ is again finite. 
Let $\vartheta$ be the maximum of all these corresponding $\vartheta_L$. 
Consequently, 
\zgl{\norm{P_{l_1} P_{l_2} \cdots P_{l_N}} \leq \vartheta}
for all $N$ and
for all sequences $l_1,\ldots,l_N$ with $\bigcap_{k=1}^N H_{l_k} = 0$.
Of course, $\vartheta < 1$.
\item
Let now $(j_k)$ be a sequence as given in the assumptions.
Since $I$ is full, there exists a strictly increasing sequence $(N_q)_{q\in\N}$ 
of natural numbers with $N_0 = 0$, 
such that 
\zgl{H_{j_{N_q+1}} \cap \ldots \cap H_{j_{N_{q+1}}} = 0}
for all $q\in\N$. 
By the preceding step 
we have $\norm{P_{j_{N_q+1}} \cdots P_{j_{N_{q+1}}}} \leq \vartheta$ 
for all $q\in\N$.
\item
Setting $A_N := \prod_{k=1}^N P_{j_k}$, we get for $Q \in \N_+$
\zgl{
\bignorm{A_{N_Q}}  
   =  \Bignorm{\prod_{q=0}^{Q-1} \:\: \prod_{s=N_q+1}^{N_{q+1}} P_s} 
 \leq \prod_{q=0}^{Q-1} \:\: \Bignorm{\prod_{s=N_q+1}^{N_{q+1}} P_s}
 \leq \prod_{q=0}^{Q-1} \vartheta 
   =  \vartheta^{Q}.
}
Consequently, $\norm{A_{N_Q}} \gegen 0$ for $Q \gegen \infty$.
Since $\norm{A_{N+1}} = \norm{A_{N} P_{j_{N+1}}} \leq \norm{A_{N}}$,
i.e., since the sequence $\norm{A_{N}}$ is non-decreasing,
we have $\norm{A_N} \gegen 0$ for $N \gegen \infty$.
\eunum
Let now $H_0 \neq 0$. Denote by $H'_i$ the orthogonal complement
of $H_0$ in $H_i$ and by $P'_i$ the corresponding projector.
Using $P_i = P_0 + P'_i$ and $P_0 P'_i = P'_i P_0 = 0$ for all $i$, we get
$\prod_{k=1}^N P_{j_k} = P_0 + \prod_{k=1}^N P'_{j_k}$ for all $N$.
By $\bigcap_{i\in I} H'_i = 0$ we have
$\prod_{k=1}^N P'_{j_k} \gegen 0$ and thus finally
$\prod_{k=1}^N P_{j_k} \gegen P_0$ for $N \gegen \infty$.

The proof of $\prod_{k=N}^1 P_{j_k} \gegen P_0$ is now clear.
\qed
\epf

\bprop
\label{prop:absch_altprod}
Let $H$ be some Hilbert space, $N \in \N$ and $\varepsilon > 0$.
Moreover, let $A$, $A_i$ and $B_i$ 
be linear continuous operators on $H$,
such that for all $i = 1,\ldots,N$
\bunum
\item
$\norm{A_i - A} \leq (1+\varepsilon)^{{2^{-i}}} - 1$ and
\item
$\norm{B_i} \leq 1$.
\eunum
If additionally $\norm{A} = 1$, then we have 
\zgl{
\Bignorm{\prod_{i=1}^N A_i B_i - \prod_{i=1}^N A B_i} < \varepsilon
\breitrel{\text{ and }}
\Bignorm{\prod_{i=1}^N B_i A_i - \prod_{i=1}^N B_i A} < \varepsilon.
}
\eprop
\bpf
We have 
\bglklein
          \bignorm{\prod_{i=1}^N A_i B_i - \prod_{i=1}^N A B_i} 
 &  =   & \bignorm{\prod_{i=1}^N (A + [A_i - A]) B_i - \prod_{i=1}^N A B_i} \\[1ex]
 & \leq & \prod_{i=1}^N \bigl(\norm{A B_i} + \norm{(A_i - A) B_i} \bigr) -
          \prod_{i=1}^N \norm{A B_i} \\[1ex]
 & \leq & \prod_{i=1}^N \bigl(\norm{A} + \norm{A_i - A} \bigr) -
          \prod_{i=1}^N \norm{A} \\[1ex]
 & \leq & \prod_{i=1}^N \bigl(1 + (1+\varepsilon)^{{2^{-i}}} -1 \bigr) -
          \prod_{i=1}^N 1 \\[1ex]
 &  =   & (1+\varepsilon)^{\sum_{i=1}^N {2^{-i}}} - 1 \\[1ex]
 &  <   & \varepsilon.
\eglklein
The proof for the opposite factor ordering is completely analogous.
\qed
\epf
Finally, we consider the special case of two projectors.
\blem
\label{lem:two_proj}
Let $P_1$ and $P_2$ be orthogonal projections on 
some Hilbert space $H$ and let
$P_0$ be the orthogonal projection from $H$ onto $P_1 H \cap P_2 H$.

Then we have for every $n \in \N_+$ 
\zgl{(P_1 P_2)^n = P_0 \breitrel\impliz P_1 P_2 = P_0.}
\elem
\bpf
\bunum
\item
Assume first $P_0 = 0$.

Since $P_1$ and $P_2$ are hermitian (i.e., in the real case, 
they equal their respective transposes), $(P_1 P_2)^{m} P_1$ is hermitian
for $m\in\N$.
Since $\norm{A^2} = \norm{A}^2$ for all hermitian operators $A$, we have
\zgl{
\norm{(P_1 P_2)^{2m} P_1} 
 = \norm{(P_1 P_2)^{m} P_1 (P_1 P_2)^{m} P_1}
 = \norm{(P_1 P_2)^{m} P_1}^2,
}
hence for all $s\in\N$
\zgl{
\norm{(P_1 P_2)^{2^s} P_1} 
 = \norm{P_1 P_2 P_1}^{2^s}.
}
Choosing some $s$ with $n \leq 2^s$, we get  
$P_1 P_2 P_1 = 0$ from $(P_1 P_2)^n = 0$.

Therefore,
$\skalprod{P_2 P_1 x}{P_2 P_1 x} = \skalprod{x}{P_1 P_2 P_1 x} = 0$  
for all $x\in H$, hence $P_2 P_1 = 0$ which implies $P_1 P_2 = 0$.
\item
Let now $P_0$ be arbitrary. 

Let $P'_i$ for $i = 1,2$ be the orthogonal projector from $H$ onto the 
orthogonal complement of $P_0 H$ in $P_i H$.
By $P_i = P_0 + P'_i$ and $P_0 P'_i = P'_i P_0 = 0$, we get
$P_0 = (P_1 P_2)^n = P_0 + (P'_1 P'_2)^n$, hence $(P'_1 P'_2)^n = 0$.
As shown above, $P'_1 P'_2 = 0$, thus $P_1 P_2 = P_0 + P'_1 P'_2 = P_0$.
\qed
\eunum
\epf

\section{Integrals of Operator Products}
\blem
\label{lem:prod_int}
Let $\gc^{(i)}$, $i=1,\ldots,k$, be finite tuples of edges
and let $\hyph^{(i)}$ for every $i=1,\ldots,k$ be some hyph with 
$\gc^{(i)} \leq \hyph^{(i)}$,
such that 
\bunum
\item
$\hyph^{(i)} \cap \hyph^{(j)} = \leeremenge$ for all $i \neq j$ and
\item
$\bigcup_i \hyph^{(i)}$ is a hyph.
\eunum
Then we have for all continuous 
$f_i : \LG^{\elanz\gc^{(i)}} \nach \C$
\bgl
 \int_\Ab \prod_i \bigl(f_i \circ \pi_{\gc^{(i)}}\bigr) \: \dd\mu_0
& = &   
     \prod_i \int_\Ab f_i \circ \pi_{\gc^{(i)}}\: \dd\mu_0.
\egl
\elem
\bpf
Define $\hyph := \bigcup_i \hyph^{(i)}$. Due to 
$\gc^{(i)} \leq \hyph^{(i)} \leq \hyph$ we have 
\bgl
 \int_\Ab \prod_i \bigl(f_i \circ \pi_{\gc^{(i)}}\bigr) \: \dd\mu_0
  & = & \int_\Ab \prod_i \Bigl(
          \bigl[\bigl(f_i \circ \pi_{\gc^{(i)}}^{\hyph^{(i)}}\bigr) 
                  \circ \pi_{\hyph^{(i)}}^\hyph \bigr] \circ \pi_\hyph 
          \Bigl) \: \dd\mu_0 \\
  & = & \int_{\LG^{\elanz{\hyph}}} \prod_i 
          \bigl[\bigl(f_i \circ \pi_{\gc^{(i)}}^{\hyph^{(i)}}\bigr) 
                  \circ \pi_{\hyph^{(i)}}^\hyph \bigr] \: \dd\mu_\Haar \\
  & = & \prod_i \int_{\LG^{\elanz{\hyph^{(i)}}}} 
          f_i \circ \pi_{\gc^{(i)}}^{\hyph^{(i)}}
                  \: \dd\mu_\Haar \\
  &   & \erl{$\hyph$ is the disjoint union of the $\hyph^{(i)}$.} \\
  & = & \prod_i \int_\Ab
          f_i \circ \pi_{\gc^{(i)}}^{\hyph^{(i)}} \circ \pi_{\hyph^{(i)}}
                  \: \dd\mu_0 \\
  & = & \prod_i \int_\Ab f_i \circ \pi_{\gc^{(i)}}\: \dd\mu_0.
\egl
\qed
\epf

\bcorr
\label{corr:prod_int_op}
Let finitely many $\tau_i \in [0,1]$ 
with $0 = \tau_0 < \tau_1 < \ldots < \tau_N = 1$ be given.
Let $\gc$ be an $n$-tuple of edges and 
define $\gc^{(i)} := \gc\einschr{[\tau_{i-1},\tau_{i}]}$.
Assume, moreover, that the reductions $\hyph^{(i)} := \redukt(\gc^{(i)})$ 
have the following two properties: 
\bunum
\item
$\hyph^{(i)} \cap \hyph^{(j)} = \leeremenge$ for all $i \neq j$ and
\item
$\bigcup_i \hyph^{(i)}$ is a hyph.
\eunum
Let now $X$ be a finite-dimensional Hilbert space and let
$F : \Ab \nach \End\:X$ be some function. Equip
$\End\:X$ with the standard operator norm induced
by the norm on $X$.
Assume finally, that there are continuous functions 
$F_i : \LG^n \nach \End\:X$, such that 
$F = \prod_i \bigl(F_i \circ \pi_{\gc^{(i)}}\bigr)$.

Then 
\bgl
 \int_\Ab F \: \dd\mu_0
  & = & \prod_i \int_\Ab F_i \circ \pi_{\gc^{(i)}} \: \dd\mu_0.
\egl
\ecorr
\bpf
Using Lemma \ref{lem:prod_int} we have for all indices $k,l$
\bgl
\Bigl( \int_\Ab F \: \dd\mu_0 \Bigr)^k_l
  & = & \int_\Ab \Bigl(\prod_i F_i \circ \pi_{\gc^{(i)}}\Bigr)^k_l
            \: \dd\mu_0 \\
  & = & \krd_{j_0}^{k} \krd^{j_N}_{l} \:
        \int_\Ab \prod_i (F_i)^{j_{i-1}}_{j_{i}} \circ \pi_{\gc^{(i)}}
            \: \dd\mu_0 \\
  & = & \krd_{j_0}^{k} \krd^{j_N}_{l} \: \prod_i 
        \int_\Ab (F_i)^{j_{i-1}}_{j_{i}} \circ \pi_{\gc^{(i)}}
            \: \dd\mu_0 \\
  & = & \Bigl(\prod_i \int_\Ab F_i \circ \pi_{\gc^{(i)}}
            \: \dd\mu_0 \Bigr)^k_l.
\egl
Note that the independence of $\bigcup_i \hyph^{(i)}$ 
implies that of every $\hyph^{(i)}$.
\qed
\epf

\section{$SU(2)$ Integral Formulae}
The basic formula \cite{Cr} we will exploit below is
\bgl
       6 \skalprod{g{}^{\mu_1}_{\nu_1} g{}^{\mu_2}_{\nu_2}}%
                {g{}^{\rho_1}_{\sigma_1} g{}^{\rho_2}_{\sigma_2}}_\Haar
 & = & 
          2 \bigl(
                 \krd^{\mu_1 \rho_1} \krd_{\nu_1 \sigma_1} \krd^{\mu_2 \rho_2} \krd_{\nu_2 \sigma_2} +  
                 \krd^{\mu_1 \rho_2} \krd_{\nu_1 \sigma_2} \krd^{\mu_2 \rho_1} \krd_{\nu_2 \sigma_1}
          \bigr) 
\\ & & \phantom{\inv 6 \Bigl( 2}
         {} -   
          \bigl(
                 \krd^{\mu_1 \rho_1} \krd_{\nu_1 \sigma_2} \krd^{\mu_2 \rho_2} \krd_{\nu_2 \sigma_1} +  
                 \krd^{\mu_1 \rho_2} \krd_{\nu_1 \sigma_1} \krd^{\mu_2 \rho_1} \krd_{\nu_2 \sigma_2}
          \bigr)
.
\egl
\noindent
Here, $g^\mu_\nu$, as usual, denotes some matrix function on $SU(2)$.
Set
$S := 6 \skalprod{g{}^{\mu_1}_{\nu_1} g{}^{\mu_2}_{\nu_2}}%
            {g{}^{\rho_1}_{\sigma_1} g{}^{\rho_2}_{\sigma_2}}_\Haar$.
\blem
\label{lem:skalp1}
$6 \skalprod{g{}^{\mu_1}_{\nu_1} g{}^{\mu_2}_{\nu_2}}%
                {g{}^{\rho_1}_{\sigma_1} g{}^{\rho_2}_{\sigma_2}}_\Haar \neq 0$
iff 
$\mu_1 + \mu_2 = \rho_1 + \rho_2$
and
$\nu_1 + \nu_2 = \sigma_1 + \sigma_2$.
\elem
\bpf
Observe first that $S = 0$ iff either
both brackets are zero or the first equals $1$ and the second equals $2$.
However, if the second were $2$, then $\mu_1 = \mu_2 = \rho_1 = \rho_2$
and $\nu_1 = \nu_2 = \sigma_1 = \sigma_2$, hence the first bracket
were $2$ implying $S =2$. 
Consequently, $S = 0$ iff both brackets are zero. 
By positivity, $S = 0$
iff each of the four Kronecker products vanishes.
Hence, $S = 0$ iff 
\bgl
0 & = &   \krd^{\mu_1 \rho_1} \krd_{\nu_1 \sigma_1} \krd^{\mu_2 \rho_2} \krd_{\nu_2 \sigma_2} 
        + \krd^{\mu_1 \rho_2} \krd_{\nu_1 \sigma_2} \krd^{\mu_2 \rho_1} \krd_{\nu_2 \sigma_1} 
\\ & & 
        {}+ \krd^{\mu_1 \rho_1} \krd_{\nu_1 \sigma_2} \krd^{\mu_2 \rho_2} \krd_{\nu_2 \sigma_1}   
        + \krd^{\mu_1 \rho_2} \krd_{\nu_1 \sigma_1} \krd^{\mu_2 \rho_1} \krd_{\nu_2 \sigma_2} \\
  & = &   \bigl(
                \krd^{\mu_1 \rho_1} \krd^{\mu_2 \rho_2} 
              + \krd^{\mu_1 \rho_2} \krd^{\mu_2 \rho_1}
          \bigr)
          \bigl(
                \krd_{\nu_1 \sigma_1} \krd_{\nu_2 \sigma_2} 
              + \krd_{\nu_1 \sigma_2} \krd_{\nu_2 \sigma_1}   
          \bigr).
\egl

The assertion can now be verified immediately.
\qed
\epf

\blem
\label{lem:skalp2}
We have 
\zgl{
6 \skalprod{g{}^{\mu_1}_{\nu_1} g{}^{\mu_2}_{\nu_2}}%
           {g{}^{\mu_1}_{\nu_1} g{}^{\mu_2}_{\nu_2}}_\Haar
 = \begin{cases}
     2 & \text{ iff $\mu_1 + \mu_2 + \nu_1 + \nu_2 \kong_2 0$ } \\
     1 & \text{ iff $\mu_1 + \mu_2 + \nu_1 + \nu_2 \kong_2 1$ } \\
   \end{cases}.
}
\elem
\bpf
We have 
\bgl
       6 \skalprod{g{}^{\mu_1}_{\nu_1} g{}^{\mu_2}_{\nu_2}}%
                {g{}^{\mu_1}_{\nu_1} g{}^{\mu_2}_{\nu_2}}_\Haar
 & = & 
          2 \bigl(
                 \krd^{\mu_1 \mu_1} \krd_{\nu_1 \nu_1} \krd^{\mu_2 \mu_2} \krd_{\nu_2 \nu_2} +  
                 \krd^{\mu_1 \mu_2} \krd_{\nu_1 \nu_2} \krd^{\mu_2 \mu_1} \krd_{\nu_2 \nu_1}
          \bigr) 
\\ & & \phantom{6 \Bigl( 2}
         {} -   
          \bigl(
                 \krd^{\mu_1 \mu_1} \krd_{\nu_1 \nu_2} \krd^{\mu_2 \mu_2} \krd_{\nu_2 \nu_1} +  
                 \krd^{\mu_1 \mu_2} \krd_{\nu_1 \nu_1} \krd^{\mu_2 \mu_1} \krd_{\nu_2 \nu_2}
          \bigr) \\
 & = & 
          2 \bigl(1 + \krd^{\mu_1 \mu_2} \krd_{\nu_1 \nu_2} \bigr) 
          - \bigl(\krd_{\nu_1 \nu_2} + \krd^{\mu_1 \mu_2} \bigr)
.
\egl
For $\mu_1 = \mu_2$, we get
$6 \skalprod{g{}^{\mu_1}_{\nu_1} g{}^{\mu_2}_{\nu_2}}%
            {g{}^{\mu_1}_{\nu_1} g{}^{\mu_2}_{\nu_2}}_\Haar
 = 1 + \krd_{\nu_1 \nu_2}$, implying the assertion.
Analogously, for $\mu_1 \neq \mu_2$, we have 
$6 \skalprod{g{}^{\mu_1}_{\nu_1} g{}^{\mu_2}_{\nu_2}}%
            {g{}^{\mu_1}_{\nu_1} g{}^{\mu_2}_{\nu_2}}_\Haar
 = 2 - \krd_{\nu_1 \nu_2}$, again implying the assertion.
\qed
\epf

\blem
\label{lem:skalp3}
$6 \skalprod{g{}^{1}_{2} g{}^{\mu}_{1}} {g{}^{1}_{1} g{}^{\mu}_{2}}_\Haar
   = (-1)^{\mu+1}$ for all $\mu$.
\elem
\bpf
The assertion follows from
\bgl
        6 \skalprod{g{}^{1}_{2} g{}^{\mu}_{1}}{g{}^{1}_{1} g{}^{\mu}_{2}}_\Haar
 & = & 
        2 \bigl(
                 \krd^{1 1} \krd_{2 1} \krd^{\mu \mu} \krd_{1 2} +  
                 \krd^{1 \mu} \krd_{2 2} \krd^{\mu 1} \krd_{1 1}
          \bigr) 
\\ & & \phantom{\inv 6 \Bigl( 2}
         {} -   
          \bigl(
                 \krd^{1 1} \krd_{2 2} \krd^{\mu \mu} \krd_{1 1} +  
                 \krd^{1 \mu} \krd_{2 1} \krd^{\mu 1} \krd_{1 2}
          \bigr) \\
 & = & 
        2 \krd^{1 \mu} - 1. 
\egl
\qed
\epf



\begin{thebibliography}{10}

\bibitem{a28}
{Abhay Ashtekar and Jerzy Lewandowski: Differential geometry on the space of
  connections via graphs and projective limits. {\it J. Geom. Phys.} {\bf 17}
  (1995) {191--230}. {\sf e-print:\ hep-th/9412073}.}

\bibitem{a30}
{Abhay Ashtekar and Jerzy Lewandowski: Projective techniques and functional
  integration for gauge theories. {\it J. Math. Phys.} {\bf 36} (1995)
  {2170--2191}. {\sf e-print:\ gr-qc/9411046}.}

\bibitem{a48}
{Abhay Ashtekar and Jerzy Lewandowski: Representation theory of analytic
  holonomy {$C^*$} algebras. In: {\it Knots and Quantum Gravity} (Riverside,
  CA, 1993), edited by John C. Baez, pp. 21--61, Oxford Lecture Series in
  Mathematics and its Applications~1 (Oxford University Press, Oxford, 1994).
  {\sf e-print:\ gr-qc/9311010}.}

\bibitem{d17}
{John C. Baez and Stephen Sawin: Diffeomorphism-invariant spin network states.
  {\it J. Funct. Anal.} {\bf 158} (1998) {253--266}. {\sf e-print:\
  q-alg/9708005}.}

\bibitem{d3}
{John C. Baez and Stephen Sawin: Functional integration on spaces of
  connections. {\it J. Funct. Anal.} {\bf 150} (1997) {1--26}. {\sf e-print:\
  q-alg/9507023}.}

\bibitem{Cr}
{Michael Creutz: {\it Quarks, Gluons and Lattices}. Cambridge University Press,
  New York, 1983.}

\bibitem{paper3}
{Christian Fleischhack: Hyphs and the Ashtekar-Lewandowski Measure. {\it J.
  Geom. Phys.} {\bf 45} (2003) {231--251}. {\sf e-print:\ math-ph/0001007}.}

\bibitem{diss}
{Christian Fleischhack: Mathematische und physikalische Aspekte
  verallgemeinerter Eichfeldtheorien im Ashtekarprogramm (Dissertation).
  Universit{\"a}t Leipzig, 2001.}

\bibitem{paper13}
{Christian Fleischhack: Parallel Transports in Webs. 
  \mbox{}MIS-Preprint 31/2003, CGPG-03/3-6. {\sf e-print:\ math-ph/0304001}.}

\bibitem{paper2+4}
{Christian Fleischhack: Stratification of the Generalized Gauge Orbit Space.
  {\it Commun. Math. Phys.} {\bf 214} (2000) {607--649}. {\sf e-print:\
  math-ph/0001006, math-ph/0001008}.}

\bibitem{e46}
{Jerzy Lewandowski and Thomas Thiemann: Diffeomorphism invariant quantum field
  theories of connections in terms of webs. {\it Class. Quant. Grav.} {\bf 16}
  (1999) {2299--2322}. {\sf e-print:\ gr-qc/9901015}.}

\bibitem{m1}
{{\textcyrm{Milan}} {\textcyrm{Prager}}: {\textcyrm{Ob odnom principe
  s\mbox{}hodimosti v prostranstve Gil\cprime berta}}. {\it
  {\textcyrm{Cheho\-slov. mat. zh.}} (Czechoslovak. J. Math.)} {\bf 10} (1960)
  {271--282}.}

\end{thebibliography}
\end{document}